\documentclass[aps,prb,twocolumn,superscriptaddress,floatfix]{revtex4-2}

\usepackage{amsmath,amssymb,bm,mathtools}
\usepackage{graphicx}
\usepackage{xcolor}
\usepackage{upgreek}
\usepackage{microtype}
\usepackage[
    colorlinks=true,
    linkcolor=blue,
    citecolor=teal,
    urlcolor=magenta,
    filecolor=magenta
]{hyperref}
\usepackage{float}
\usepackage{pgfplots}
\pgfplotsset{compat=1.18}
\usepgfplotslibrary{fillbetween}
\hypersetup{colorlinks=true,linkcolor=blue,citecolor=blue,urlcolor=blue}
\allowdisplaybreaks[4]

\begin{document}

\title{Field theory of competing phases in the interacting Kitaev chain}

\author{Pradip Kattel}
\email{pradip.kattel@unige.ch}

\author{Thierry Giamarchi}
\affiliation{Department of Quantum Matter Physics, University of Geneva, Quai Ernest-Ansermet 24, 1211 Geneva, Switzerland}

\date{\today}

\begin{abstract}
We study the interacting Kitaev chain with nearest-neighbor interactions and finite chemical potential using a combination of
bosonization and density-matrix renormalization-group calculations. Focusing away from special exactly solvable lattice limits,  we develop a continuum field-theoretic description of the generic finite-density interacting regime, where doping nonperturbatively modifies the competition between superconducting pairing and commensurate charge ordering. We show that the phase diagram consists of an intermediate incommensurate Luttinger liquid between the topological superconducting and commensurate charge-density-wave Mott phases, as well as a trivial band-insulating regime at large chemical potential. The commensurate--incommensurate and incommensurate--topological-superconducting phase boundaries are governed by Pokrovsky--Talapov and Berezinskii--Kosterlitz--Thouless criticality, respectively. We analyze the interacting topological superconducting regime and its boundary excitations. At the semiclassical level, and explicitly in the free-fermion boundary theory, the topological phase supports exponentially localized Majorana edge modes at interfaces between topological and trivial regions. We also construct a dual continuum fermionic representation in which this topological bulk sector is described instead by conventional symmetry breaking. This gives a continuum version of the SPT--SSB correspondence between different fermionic representations. Overall, our results provide an analytical description of one-dimensional interacting topological superconductors, from the competition between topology and commensurability and the resulting quantum criticality to the dual symmetry-breaking description of the topological phase.
\end{abstract}

\maketitle
\section{Introduction}\label{intro}

The Kitaev chain is one of the simplest models of a one-dimensional topological superconductor. Its topological phase supports exponentially localized Majorana edge modes and has a nontrivial fermion-parity structure in the ground state~\cite{kitaev2001unpaired}. The noninteracting model can be solved exactly in terms of decoupled Majorana fermions. Interactions change this picture by introducing competing ordering tendencies and making commensurability important, leading to interaction-driven quantum phase transitions~\cite{gangadharaiah2011majorana}.

These effects have been studied analytically and numerically in the interacting Kitaev chain~\cite{stoudenmire2011interaction,miao2017exact,miao2018majorana,herviou2016phase,nunziante2024topological,mahyaeh2020study}. Through a Jordan-Wigner transformation, the model maps onto an anisotropic spin chain in a magnetic field~\cite{lieb1961two,fradkin2013field}. In general, the resulting XYZ chain with a longitudinal field is nonintegrable, although several important exactly solvable limits are known. Previous works have explored a variety of strongly interacting regimes of this model. In particular, Ref.~\cite{chepiga2023eight} mapped out the phase diagram using exact results for the XYZ chain together with extensive DMRG. Away from particle-hole symmetry, an incommensurate Luttinger liquid separates the period-two charge-ordered phase from the $\mathbb{Z}_2$ topological phase. The transitions are Pokrovsky--Talapov and Kosterlitz--Thouless, respectively. At particle-hole symmetry, the two phases meet directly along the eight-vertex critical line. Another important direction was developed in  Ref.~\cite{katsura2015exact}, where exact frustration-free ground states and interacting generalizations of Majorana edge operators were constructed on special solvable lines of the interacting Kitaev chain. The boundary physics has also been approached directly in the continuum. Open-boundary bosonization has been used to describe Majorana edge modes in one-dimensional topological superconductors~\cite{gangadharaiah2011majorana,chua2020majorana}, with related methods applied to boundary states in the interacting SSH chain~\cite{jin2023bosonization}.

Experimental signatures of one-dimensional topological superconductivity have been studied in semiconductor--superconductor hybrid nanowires~\cite{mourik2012signatures,das2012zero} and magnetic atom chains on superconducting substrates~\cite{nadj2014observation,pawlak2016probing}. Related realizations have also been proposed in synthetic quantum systems, including ultracold atomic gases~\cite{jiang2011majorana}. In these platforms, the chemical potential can be tuned by electrostatic gates or by changing the particle density. Moreover, interactions can become important in the effectively one-dimensional regime. As a result, experiments naturally probe regimes away from particle-hole symmetry and exactly solvable limits, where interactions and commensurability compete with topological superconductivity. Understanding this finite-density interacting regime is essential for interpreting experimental signatures of topological superconductivity and identifying competing ordered phases that may obscure or replace Majorana physics.

This motivates us to determine the phases that occur in the interacting finite-density regime and the quantum phase transitions between them. A central question is how topological superconductivity competes with interaction-driven charge order as the system is doped away from commensurability. At low energies, this can be studied within a bosonized description of the interacting Kitaev chain~\cite{haldane1981luttinger,giamarchi2003quantum,gogolin2004bosonization}. Superconducting pairing tends to pin the phase field $\theta$ and open a topological superconducting gap. Repulsive interactions, on the other hand, generate Umklapp processes that tend to pin the density field $\phi$, leading to a commensurate charge-density-wave Mott phase. Since $\phi$ and $\theta$ are canonically conjugate, the competition between these two locking mechanisms is intrinsically nonperturbative. A finite chemical potential frustrates the locking of $\phi$. Once it becomes large enough, sine-Gordon solitons proliferate and drive a Pokrovsky--Talapov transition into an intermediate incommensurate Luttinger liquid~\cite{pokrovsky1979ground}.

In this work, we combine bosonization and renormalization-group analysis with density-matrix renormalization-group calculations to determine the phase diagram of the interacting Kitaev chain. Doping the commensurate charge-density-wave Mott phase produces an incommensurate Luttinger liquid before the system enters the topological superconducting phase. The two transitions are of Pokrovsky--Talapov and Berezinskii--Kosterlitz--Thouless type. At larger chemical potential, the topological superconducting phase undergoes an Ising transition to a trivial band insulator, with a single massless Majorana mode and $z=1$, $c=1/2$. We also study the Majorana edge modes of the topological phase. In the absence of interactions, the open-chain theory can be solved explicitly and gives exponentially localized Majorana modes at the two ends. For finite interactions, bosonization retains the fermion-parity structure of the open chain and is consistent with a quasi-local dressing of these modes. We finally construct a dual fermionic representation in which the same low-energy bulk theory exhibits conventional discrete symmetry breaking, thereby providing a continuum realization of the SPT--SSB correspondence.

The paper is organized as follows.  In Sec.~\ref{model}, we introduce the interacting Kitaev chain and derive its low-energy bosonized description. In Sec.~\ref{sec:RG}, we analyze the scaling dimensions of the competing superconducting and Umklapp perturbations and derive the corresponding renormalization-group flows. In Sec.~\ref{sec:Phase_transtions}, we discuss the resulting phases and quantum phase transitions and compare the field-theoretic predictions with DMRG calculations. In Sec.~\ref{sec:field-theory-topological}, we turn from the bulk phase diagram to the topological phase itself, examining its boundary physics before developing a dual symmetry-breaking description of the same gapped bulk. We show how Majorana edge modes arise in the continuum theory and how fermion parity organizes the Hilbert space of an open chain. Finally, in Sec.~\ref{conclusion}, we summarize our results and conclusions.

\section{Model}\label{model}

We consider the interacting Kitaev chain described by the following spinless fermionic Hamiltonian
\begin{align}
H&=\sum_j -t\left(c_j^\dagger c_{j+1}+c_{j+1}^\dagger c_j\right)+\Delta\left(c_j c_{j+1}+c_{j+1}^\dagger c_j^\dagger\right)\nonumber\\
&\qquad\quad+U\left(2n_j-1\right)\left(2n_{j+1}-1\right)
-\mu\left(n_j-\frac12\right).
\label{eq:lattice_hamiltonian}
\end{align}
where $n_j=c_j^\dagger c_j$. Here $t$ denotes the nearest-neighbor hopping amplitude, $\Delta$ is the $p$-wave superconducting pairing strength, $U$ is the nearest-neighbor density interaction, and $\mu$ is the chemical potential. Although the pairing term explicitly breaks the global $U(1)$ particle-number symmetry, the Hamiltonian preserves fermion parity
\begin{equation}
P=(-1)^F,\quad F=\sum_j n_j,
\end{equation}
since the fermion number changes only in pairs. In the absence of interactions, Eq.~\eqref{eq:lattice_hamiltonian} reduces to the Kitaev chain and supports a topological superconducting phase with exponentially localized Majorana edge modes for $|\mu|<2t$.

Our goal is to develop a low-energy description of the competing bulk phases and interacting boundary excitations of Eq.~\eqref{eq:lattice_hamiltonian}. We first characterize the bulk phase diagram using bosonization and renormalization-group methods. We then turn to the topological regime, where a fermionic mass-domain-wall description and finite-size bosonization are used to study the boundary modes and fermion-parity structure, while the interacting sine-Gordon theory extends the analysis to finite interactions.

To derive the continuum theory, we expand the lattice fermion operator around the two Fermi points,
\begin{equation}
c_j\rightarrow\sqrt{a_0} \psi(x),\quad x=j a_0,
\end{equation}
with
\begin{equation}
\psi(x)=e^{ik_F^0x}\psi_R(x)+e^{-ik_F^0x}\psi_L(x),
\label{eq:continuum_expansion}
\end{equation}
where $a_0$ is the lattice spacing and $\psi_R(x)$ and $\psi_L(x)$ denote slowly varying right- and left-moving fermion fields.

For the low-energy description of commensurability, we expand about the half-filled reference state with Fermi momentum $k_F^0=\frac{\pi}{2a_0}$. The chemical potential $\mu$ is then treated as the field conjugate to the deviation from this commensurate reference density, rather than being absorbed into a shifted Fermi momentum. We employ Abelian bosonization \cite{giamarchi2003quantum} and introduce dual bosonic fields $\phi(x)$ and $\theta(x)$ satisfying 
\begin{equation}
[\phi(x),\partial_y\theta(y)]=i\pi\delta(x-y).
\label{eq:commutator}
\end{equation}
The slowly varying chiral fermion fields are represented as vertex operators,
\begin{equation}
\psi_r(x)=\frac{1}{\sqrt{2\pi a_0}}\exp\left[-i\left(
r\phi(x)-\theta(x)\right)\right],\quad r=\pm1,
\label{eq:bosonization_dictionary}
\end{equation}
where $r=+1$ and $r=-1$ correspond to right- and left-moving fermions, respectively.

The Gaussian sector of the theory is described by the Luttinger-liquid Hamiltonian
\begin{equation}
H_{LL} = \frac{v}{2\pi} \int \mathrm{d}x \left[ K(\partial_x\theta)^2 + \frac{1}{K}(\partial_x\phi)^2
\right],
\label{eq:LL_hamiltonian}
\end{equation}
where $v$ is the renormalized velocity and $K$ is the Luttinger parameter. Repulsive density interactions reduce $K$, whereas attractive interactions increase it.

The continuum density takes the standard bosonized form
\begin{equation}
\rho(x) = \rho_0 - \frac{1}{\pi}\partial_x\phi + A\cos\left( 2k_F^0x-2\phi(x) \right) + \cdots.
\label{eq:density_bosonized}
\end{equation}
The chemical-potential term becomes
\begin{equation}
H_\mu = -\mu\int \mathrm{d}x \rho(x)
\label{eq:chemical_potential_phi}
\end{equation}
up to an additive constant proportional to the system size. Equivalently, the chemical potential couples directly to the topological charge
\begin{equation}
Q_\phi = -\frac{1}{\pi} \int \mathrm{d}x \partial_x\phi,
\label{eq:topological_charge}
\end{equation}
through $H_\mu=-\mu Q_\phi$. Here $\mu$ is measured relative to the commensurate reference state and acts as the field conjugate to the topological charge $Q_\phi$. For sufficiently small $|\mu|$ in the gapped commensurate phase, the density remains pinned despite the finite chemical potential; only when the soliton gap is overcome does a finite density of topological solitons develop.

Using Eq.~\eqref{eq:bosonization_dictionary}, the superconducting pairing term generates the perturbation
\begin{equation}
H_\Delta = \frac{v y_\Delta}{2\pi a_0^2} \int \mathrm{d}x \cos\left( 2\theta \right),
\label{eq:pairing_cosine}
\end{equation}
while the interaction-induced Umklapp process produces
\begin{equation}
H_U = \frac{v y_U}{2\pi a_0^2} \int \mathrm{d}x\,\cos(4\phi).
\label{eq:umklapp_cosine}
\end{equation}

The effect of doping away from commensurability is encoded by the chemical-potential coupling to $Q_\phi$, rather than by an independent shift of the Fermi momentum in the Umklapp operator. The low-energy theory thus takes the form
\begin{align}
H_{\mathrm{eff}} &= H_{LL} + \frac{\mu}{\pi}
\int \mathrm{d}x \partial_x\phi + \frac{v y_\Delta}{2\pi a_0^2}
\int \mathrm{d}x \cos\left( 2\theta \right) \nonumber\\
&\quad+ \frac{v y_U}{2\pi a_0^2} \int \mathrm{d}x
\cos\left( 4\phi\right),
\label{eq:effective_hamiltonian}
\end{align}
which starts for the analysis of the competing bulk phases~\cite{giamarchi1988theory,takayoshi2018topological} and interacting boundary physics discussed below.

\section{Scaling Dimensions and Renormalization Group}\label{sec:RG}

The phase diagram is controlled by the competition between superconducting pairing and commensurate charge ordering. Within the low-energy description Eq.~\eqref{eq:effective_hamiltonian}, these tendencies are encoded by the pairing perturbation $\cos(2\theta)$ and the Umklapp perturbation $\cos(4\phi)$, respectively. Their competition is governed by the scaling dimensions of the corresponding vertex operators.

Introducing the general operators
\begin{equation}
V_{n,m}(x) = e^{i(n\phi+m\theta)},
\end{equation}
the Gaussian correlations generated by Eq.~\eqref{eq:LL_hamiltonian} yield
\begin{equation}
\Delta_{n,m}  = \frac14 \left( Kn^2+\frac{m^2}{K}
\right).
\label{eq:vertex_dimension}
\end{equation}

The pairing perturbation
$ \cos(2\theta) = \frac12 \left( V_{0,2}+V_{0,-2}
\right) $
has a scaling dimension $\Delta_\Delta = \frac1K,
$ while the Umklapp perturbation
$ \cos(4\phi) = \frac12 \left( V_{4,0}+V_{-4,0}
\right) $
has a scaling dimension $\Delta_U = 4K$.

The corresponding tree-level renormalization-group equations are
\begin{equation}
\frac{dy_\Delta}{dl} = \left( 2-\frac1K \right)y_\Delta,
\label{eq:RG_pairing}
\end{equation}
and
\begin{equation}
\frac{dy_U}{dl} = \left( 2-4K \right)y_U.
\label{eq:RG_umklapp}
\end{equation}
The pairing perturbation is relevant for $K>1/2$, whereas the Umklapp perturbation becomes relevant for $K<1/2$.

Beyond tree level, the perturbations also renormalize the Luttinger parameter itself. A standard momentum-shell calculation yields
\begin{equation}
\frac{dK}{dl} = \frac12 y_\Delta^2 - 2 K^2 y_U^2.
\label{eq:RG_K}
\end{equation}
The pairing interaction increases $K$, thereby reinforcing superconducting correlations, while the Umklapp interaction decreases $K$ and favors commensurate charge order. More generally, the two perturbations $\cos(2\theta)$ and $\cos(4\phi)$ describe competing ordering tendencies in the canonically conjugate fields $\theta$ and $\phi$, respectively. Since these fields cannot be simultaneously pinned, the resulting phase is determined by the full coupled renormalization-group flow of $K$, $y_\Delta$, and $y_U$. The flow and the Hamiltonian are similar to the ones 
of the problem of spin-orbit coupling in one-dimensional interacting systems and contain a separatrix separating the basins of attraction of the superconducting and commensurate charge-ordered strong-coupling regimes~\cite{giamarchi1988theory}.

The same coefficients are consistent with the continuum duality introduced
below. Under $K\rightarrow K_{\rm d}=1/(4K)$ and
$y_U\leftrightarrow y_\Delta$, Eq.~\eqref{eq:RG_K} gives
\begin{equation}
\frac{dK_{\rm d}}{dl}=\frac12y_U^2-2K_{\rm d}^2y_\Delta^2,
\label{eq:RG_duality_check}
\end{equation}
which has the same form as the original flow.

Roughly, $K>1/2$ favors the superconducting instability where the pairing perturbation is relevant at weak coupling and tends to pin $\theta$, driving the system toward the topological superconducting phase. In the continuum description, the topological nature of this strong-coupling regime manifests itself through nontrivial boundary excitations and fermion-parity structure, which will be analyzed below using fermionization, finite-size bosonization, and the interacting sine-Gordon description. Conversely, for $K<1/2$, the Umklapp perturbation is relevant at weak coupling and tends to pin $\phi$, favoring the commensurate charge-density-wave Mott phase. The actual boundary between these two strong-coupling regimes is, however, determined by the separatrix of the coupled RG flow rather than by the bare value $K=1/2$ alone.

Away from the commensurate plateau, the chemical potential favors a finite topological charge and eventually nucleates solitons of the $\phi$ field. Once a finite soliton density develops, the resulting background gradient of $\phi$ generates an incommensurate phase mismatch in the Umklapp operator, suppressing commensurate locking at long distances. This allows a gapless incommensurate Luttinger liquid to intervene between the commensurate Mott and topological superconducting phases, as discussed below.

\section{Phases and Quantum Phase Transitions}\label{sec:Phase_transtions}

We now discuss the phases generated by the renormalization-group flows derived above and compare the field-theoretic predictions with representative density-matrix renormalization-group calculations for $t=1$ and $\Delta=0.5$. The phase diagram contains four regimes:
\begin{enumerate}
\item a topological superconducting phase,
\item a commensurate charge-density-wave Mott phase,
\item an intermediate incommensurate Luttinger liquid,
\item and a trivial band insulator at sufficiently large chemical potential.
\end{enumerate}

\subsection{Topological superconducting phase}

The topological superconducting phase is characterized by a finite bulk excitation gap together with exponentially localized boundary excitations. In the bosonized description, these properties originate from the pairing perturbation $\cos(2\theta)$, which becomes relevant for $K>\frac12$. Under the renormalization-group flow, the coupling $y_\Delta$ flows to strong coupling and dynamically generates the superconducting mass scale that gaps the bulk. The resulting phase is continuously connected to the topological phase of the noninteracting Kitaev chain, corresponding to $K=1$ in the bosonized description, and inherits its nontrivial fermion-parity structure~\cite{alicea2012new}.
\begin{figure}
    \centering    \includegraphics[width=\linewidth]{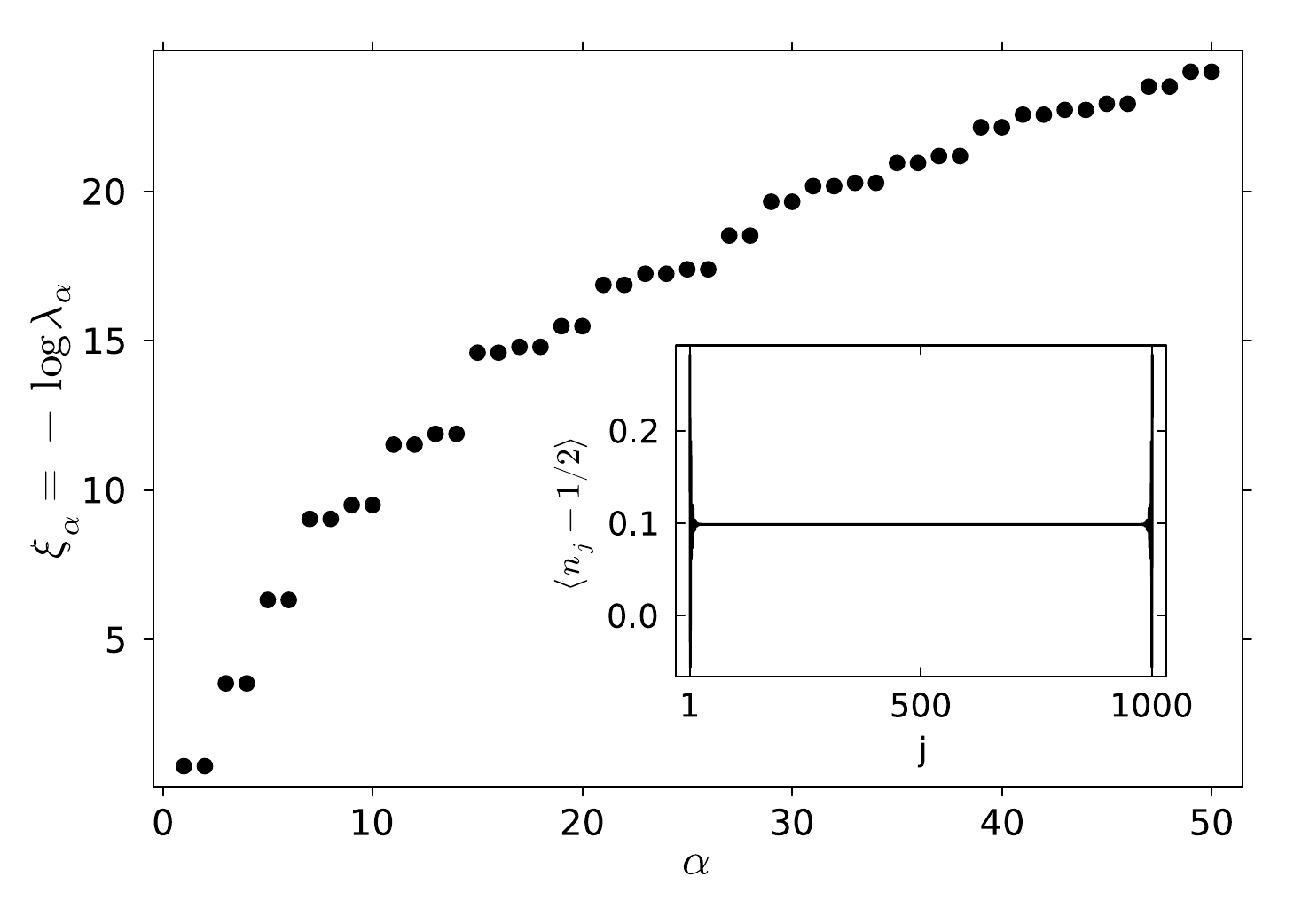}
    \caption{Lowest $50$ entanglement levels $\xi_\alpha=-\log\lambda_\alpha$, where $\lambda_\alpha$ are the eigenvalues of the reduced density matrix, for the interacting Kitaev chain in the topological superconducting regime with chemical potential $\mu=1.5$, interaction strength $U=0.5$, hopping amplitude $t=1$, and pairing strength $\Delta=0.5$. The near-double degeneracy of the low-lying entanglement spectrum is consistent with the interacting topological phase. \textit{Inset}: local density profile $\langle n_j-1/2\rangle$, where $n_j=c_j^\dagger c_j$ is the fermion density at site $j$, showing a featureless bulk with weak Friedel oscillations near the boundaries.}
    \label{fig:entanglement-spectrum}
\end{figure}

In the DMRG calculations, the topological superconducting phase is identified through a combination of bulk and entanglement signatures, including a finite many-body gap, the near-double degeneracy of the bipartite entanglement spectrum obtained from a half-chain cut, and the absence of conventional bulk density order~\cite{pollmann2010entanglement}. The entanglement spectrum is defined from the eigenvalues $\lambda_\alpha$ of the reduced density matrix $\rho_A$ through $\xi_\alpha = -\log \lambda_\alpha$. The numerical calculations were performed on open chains with $N=1000$ sites using up to $35$ DMRG sweeps with a truncation cutoff of $10^{-12}$ and convergence monitored through an energy tolerance criterion $\Delta E < 10^{-12}$. In the topological regime, the local density profile $\langle n_j - 1/2 \rangle$ remains essentially featureless throughout the bulk, with only weak boundary-induced Friedel oscillations near the edges. Representative entanglement spectra exhibit pairs of nearly degenerate levels, consistent with the persistence of the interacting topological phase away from the free-fermion limit~\cite{pollmann2010entanglement}. In Fig.~\ref{fig:entanglement-spectrum}, we show these features for a point in the topological regime with $\mu = 1.5$ and $U = 0.5$.

\subsection{Commensurability, Umklapp locking, and the Mott phase}\label{sec:mott}

At exact commensurability, the density is locked to the half-filled reference state with $k_F^0=\pi/(2a_0)$, and the Umklapp perturbation $\cos(4\phi)$ is momentum conserving. As discussed in the previous section, for sufficiently strong repulsive interactions, the Umklapp coupling flows to strong coupling, dynamically generating a Mott gap. The resulting strong-coupling phase is a fully gapped commensurate charge-density-wave insulator. For definiteness, we choose $y_U<0$, in which case the bosonic field $\phi$ becomes pinned at one of the minima of the cosine potential, $\phi=\frac{n\pi}{2}, \quad n\in\mathbb{Z}$.

The emergence of charge order follows directly from the staggered component of the bosonized density, $\rho_{\rm stag}(x)\sim\cos(2\phi)$. Under translation by one lattice spacing, $\phi\rightarrow\phi+\frac{\pi}{2}$, which exchanges adjacent minima of the cosine potential. Thus, pinning $\phi$ spontaneously breaks the discrete lattice translation symmetry and produces staggered density order,
\begin{equation}
\langle n_j\rangle = \rho_0+(-1)^j m_{\mathrm{CDW}}.
\end{equation}

\begin{figure}
    \centering
    \includegraphics[width=\linewidth]{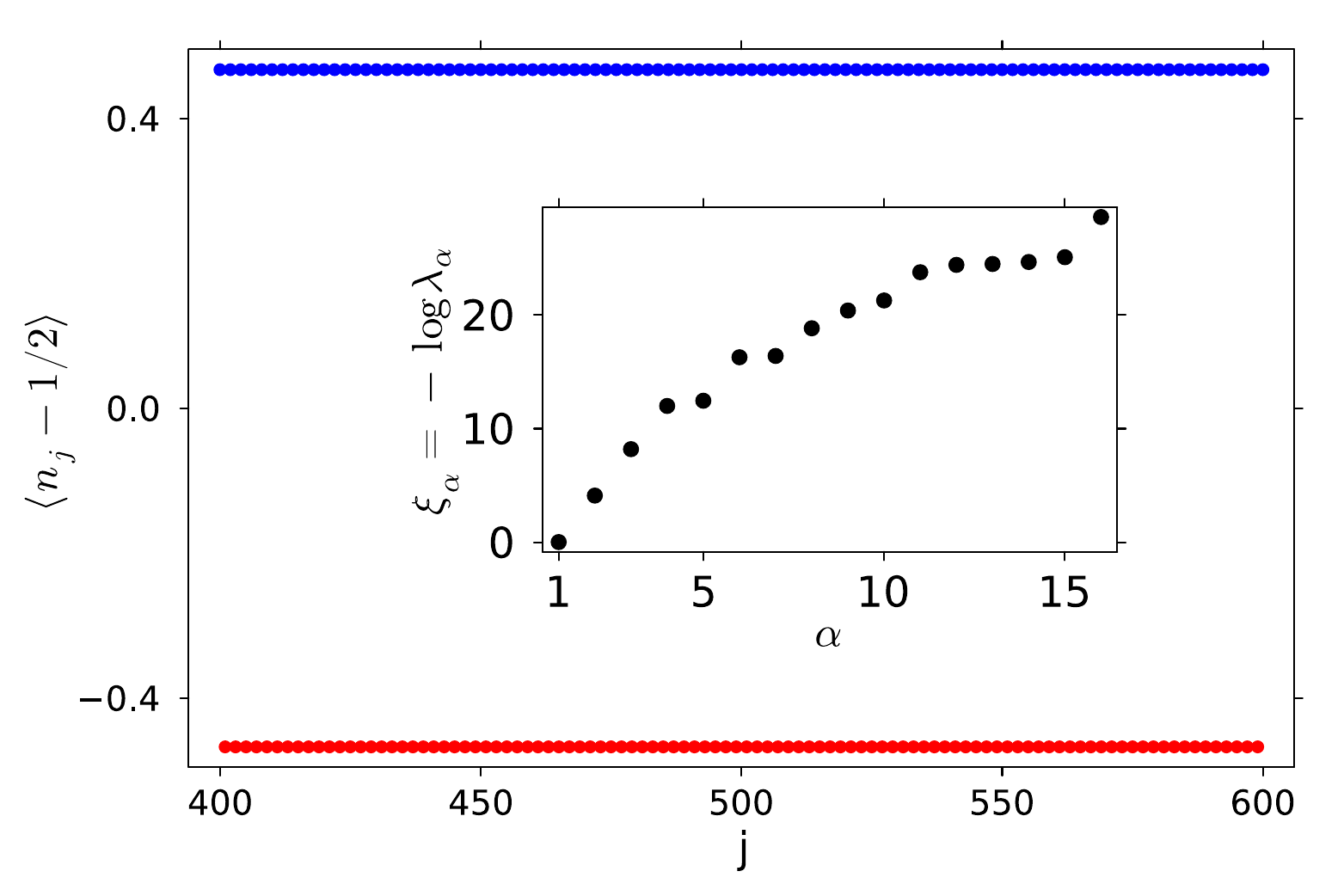}
    \caption{Representative DMRG results in the commensurate Mott phase for $\mu = 3$, $U = 2$, $t = 1$, and $\Delta = 0.5$ on an open chain with $N = 1000$ sites. \textit{Main panel}: local density profile $\langle n_j - 1/2 \rangle$ showing robust period-two charge-density-wave order in the bulk. Even and odd sublattices are plotted separately using different marker styles to emphasize the staggered CDW structure. \textit{Inset}: lowest entanglement levels $\xi_\alpha = -\log \lambda_\alpha$, which are nondegenerate, consistent with the absence of topological boundary modes in the Mott insulating phase.}
    \label{fig:cdw-n_j}
\end{figure}

Away from the particle-hole-symmetric line, the incommensurate--topological
superconducting boundary is controlled by the weak-coupling condition that
the pairing perturbation becomes marginal at $K=1/2$~\cite{giamarchi1988theory}.
This boundary is of Berezinskii--Kosterlitz--Thouless type~\cite{berezinskii1971destruction,kosterlitz1973ordering}. Approaching it
from the superconducting side, the gap has the essential singularity
\begin{equation}
M \sim D\exp\left(-\frac{A}{\sqrt{\delta}}\right),
\quad \delta\rightarrow0^+,
\label{eq:BKT_gap}
\end{equation}
where $D$ is the ultraviolet energy scale, $A$ is non-universal, and
$\delta$ measures the distance from the BKT separatrix.

At $\mu=0$ the Jordan--Wigner transformation of Eq.~\eqref{eq:lattice_hamiltonian} is the zero-field XYZ chain with
\begin{equation}
J_x=-\frac{t+\Delta}{2},\quad J_y=-\frac{t-\Delta}{2},\quad J_z=U.
\label{eq:XYZ_couplings}
\end{equation}
The XYZ chain is integrable for all couplings, and it is critical when the two couplings of largest magnitude are equal in magnitude. For $0<\Delta<t$ one has $|J_y|<|J_x|$, so the direct transition occurs at $|J_z|=|J_x|$, i.e. $J_z=-J_x$. For $\Delta\neq0$ this is an XYZ transition and is described by the eight-vertex universality class. The XXZ limit is obtained when $\Delta=0$; only in that limit does the direct transition become the usual BKT transition. The transition occurs at the exactly known coupling
\begin{equation}
U_c=\frac{t+\Delta}{2}.
\label{eq:exact_Uc}
\end{equation}

Rotating every second spin by $\pi$ about the $z$ axis reverses the signs of $J_x$ and $J_y$, so that at $U=U_c$ the chain becomes the antiferromagnetic XXZ chain $\frac{t+\Delta}{2}\sum_j(\sigma^x_j\sigma^x_{j+1}+\sigma^z_j\sigma^z_{j+1}+\cos\rho\,\sigma^y_j\sigma^y_{j+1})$ with symmetry axis along $y$ and $\cos\rho=(t-\Delta)/(t+\Delta)$. Both the staggered density $(-1)^j\sigma^z_j$ and the Kitaev order parameter $\sigma^x_j$ are in-plane spin components of this chain. The critical theory is a $c=1$ Gaussian theory with an emergent $U(1)$ symmetry about the $y$ axis. Its Luttinger parameter is $K_{\rm 8v}=\pi/[2(\pi-\rho)]$, with $\cos\rho=(t-\Delta)/(t+\Delta)$~\cite{chepiga2023eight}. The common scaling dimension $K^\ast$ of the two order parameters is related to this parameter by $K^\ast=1/(4K_{\rm 8v})$. The parameter $K_{\rm 8v}$ should not be confused with the Luttinger parameter $K$ of Eq.~\eqref{eq:effective_hamiltonian}, which is defined with respect to the fermion-number $U(1)$. At the XXZ endpoint $\Delta=0$, one has $K_{\rm 8v}=1/2$, while along the eight-vertex line at $\Delta\neq0$ one has $K_{\rm 8v}>1/2$. Thus, $K^\ast=\frac{\beta}{\nu}$. The perturbation driving the system away from the critical line, the in-plane anisotropy $\sigma^x\sigma^x-\sigma^z\sigma^z$, has scaling
dimension $4K^\ast$, which gives
\begin{equation}
\nu=\frac{1}{2-4K^\ast},\quad \beta=\nu K^\ast,
\label{eq:eightvertex_exponents}
\end{equation}
and eliminating $K^\ast$ gives the relation
\begin{equation}
4\beta=2\nu-1
\label{eq:eightvertex_relation}
\end{equation}
obeyed along the eight-vertex critical line. Matching to Baxter's solution, $K^\ast=(\pi-\rho)/2\pi$ with $\cos\rho=J_y/J_x=(t-\Delta)/(t+\Delta)$,
reproduces $\nu=\pi/2\rho$ and $\beta=(\pi-\rho)/4\rho$~\cite{baxter1971eight,baxter1985exactly,chepiga2023eight}. 

A finite chemical potential does not immediately destroy commensurability, because the density remains pinned as long as the
energy required to create a topological soliton remains positive. Once the chemical potential exceeds the soliton threshold, however, a finite density of solitons develops. Their proliferation generates an incommensurate background gradient of $\phi$ and suppresses the long-distance locking produced by the Umklapp perturbation. Hence, away from the commensurate plateau, the direct transition between the topological superconducting and Mott phases can be split by an intermediate incommensurate Luttinger liquid. The corresponding Pokrovsky--Talapov mechanism is discussed below.

The Mott phase exhibits long-range staggered density order together with a nondegenerate entanglement spectrum, consistent with the absence of topological boundary modes. In Fig.~\ref{fig:cdw-n_j}, we show DMRG results for the Mott phase at $\mu = 3$ and $U = 2$ for an open chain with $N = 1000$ sites. The local density profile exhibits nearly perfect period-two charge-density-wave order, while the entanglement spectrum no longer displays the even-fold degeneracy characteristic of the topological superconducting phase.

\subsection{Solitons and Pokrovsky--Talapov transition}

Away from half filling, the staggered density order becomes increasingly frustrated by doping. In the bosonized description, this effect arises because the chemical potential frustrates the commensurate locking of the field $\phi$. From Eq.~\eqref{eq:chemical_potential_phi}, the chemical potential couples directly to the conserved topological charge $Q_\phi$ given by Eq.\eqref{eq:topological_charge} through $H_\mu=-\mu Q_\phi$.

When the Umklapp perturbation flows to strong coupling, it generates a massive sine-Gordon phase whose elementary excitations are topological solitons interpolating between adjacent minima of the cosine potential, $\Delta\phi=\pm\frac{\pi}{2}$~\cite{coleman1975quantum,zamolodchikov1979factorized,bortolin2019quantum,suleiman2026electrostatically}. Each elementary soliton carries a fractional topological charge
$\Delta Q_\phi=\mp\frac{1}{2}$, and couples directly to the chemical potential. For $\mu>0$ the chemical potential favors $\Delta Q_\phi=+1/2$, i.e. solitons with $\Delta\phi=-\pi/2$, and conversely for $\mu<0$.

The resulting sine-Gordon theory is integrable~\cite{coleman1975quantum,zamolodchikov1979factorized} and admits an exact Bethe-Ansatz description in terms of soliton rapidities. In the commensurate Mott phase below the Pokrovsky--Talapov threshold, the ground state contains no finite density of physical solitons. Once the chemical potential exceeds the soliton gap, solitons enter the ground state and form a finite-density Bethe sea, producing the incommensurate phase.

The energy required to create a domain wall in the commensurate density pattern defines the soliton gap. The integrable sine-Gordon mass--coupling relation~\cite{bergknoff1979structure,zamolodchikov1995mass} implies the scaling form
\begin{equation}
M_s = C_{\mathrm{SG}}(K)\,D\,|y_U|^{1/(2-4K)}, \quad 0<K<\frac12,
\label{eq:Ms}
\end{equation}
where $y_U$ is the dimensionless bare Umklapp coupling appearing in Eq.~\eqref{eq:umklapp_cosine}, $D\sim v/a_0$ is the ultraviolet cutoff, and $C_{\mathrm{SG}}(K)$ is a dimensionless amplitude. Once the normalization of
the sine-Gordon theory is fixed, this amplitude can be written in terms of ratios of Gamma functions. Because the relation between the microscopic lattice parameters and the continuum coupling $y_U$ is non-universal, we
use Eq.~\eqref{eq:Ms} only for its scaling dependence.

The fixed-$K$ power-law expression in Eq.~\eqref{eq:Ms} is not uniform as $K\rightarrow\frac12^{-}$. In this limit, the Umklapp perturbation becomes marginal, and the coupled renormalization-group flow of $K$ and $y_U$ must be used instead. This produces the Berezinskii--Kosterlitz--Thouless essential singularity discussed above.

The chemical potential lowers the energy of an elementary soliton of the favored sign by $|\mu|/2$. The corresponding soliton formation energy is $E_s=M_s-\frac{|\mu|}{2}$, so that the commensurate--incommensurate transition  occurs when
\begin{equation}
\frac{|\mu_c|}{2}=M_s.
\label{eq:PT_threshold}
\end{equation}

Once the soliton gap is overcome, the system develops a finite density of solitons. In the original density-wave picture, these excitations correspond to mobile domain walls separating locally commensurate regions. Their proliferation destroys the commensurate order and produces an incommensurate phase with a continuously varying modulation wavevector. Within the Bethe Ansatz description, this corresponds to a finite density of solitons in the ground state. Near the transition, the low-density equation of state takes the universal Pokrovsky--Talapov form~\cite{iwabuchi1995commensurate,schulz1980critical}
\begin{equation} 
\frac{E-E_0}{L} = \left(M_s-\frac{|\mu|}{2}\right)n_s +\frac{\pi^2v^2}{6M_s} n_s^3, 
\label{eq:PT_eos} 
\end{equation}
where $n_s$ denotes the soliton density. Minimization yields $n_s \propto \sqrt{\frac{|\mu|}{2}-M_s}$, which is the characteristic square-root onset of the commensurate-incommensurate transition.

Since each favored soliton carries the winding $\Delta\phi=-\operatorname{sgn}(\mu)\,\frac{\pi}{2}$, a finite soliton density generates a uniform background gradient $\langle\partial_x\phi\rangle=-\operatorname{sgn}(\mu)\,\frac{\pi}{2}n_s$. Hence, writing $\phi$ for the fluctuation about this background, the Umklapp perturbation acquires an incommensurate phase mismatch,
\begin{equation}
\cos\left[4\phi-\operatorname{sgn}(\mu)\,\delta q\, x\right],
\end{equation}
with modulation wavevector
\begin{equation}
\delta q = 4|\langle\partial_x\phi\rangle|
= 2\pi n_s \propto \sqrt{\frac{|\mu|}{2}-M_s}.
\end{equation}

Near threshold, the soliton gas becomes dilute and nonrelativistic with quadratic dispersion
$E(k)\sim k^2$, corresponding to a dynamical exponent $z=2$~\cite{giamarchi1997mott,giamarchi2003quantum}.

\subsection{Incommensurate Luttinger liquid}

The incommensurate phase is characterized by gapless excitations and density modulations with a continuously varying wavevector. In the field-theoretic description, these properties originate from the finite density of solitons generated beyond the commensurate-incommensurate transition. The finite soliton density shifts the effective Fermi momentum away from the commensurate value. Consequently, the Umklapp interaction acquires a rapidly oscillating phase factor and becomes irrelevant under the renormalization-group flow. In the incommensurate regime, the pairing perturbation is also irrelevant because its scaling dimension is $1/K>2$. The low-energy theory flows back to a Gaussian Luttinger-liquid fixed point. This phase corresponds to the incommensurate regime discussed in Ref.~\cite{giamarchi1988theory,giamarchi2003quantum}, in which the departure from commensurability suppresses the Umklapp locking and leaves a gapless Luttinger liquid at low energies.

The absence of a gap implies power-law correlations characteristic of a Luttinger liquid. The elementary low-energy excitations correspond to collective density fluctuations of the soliton liquid. Both superconducting and charge-density-wave correlations decay algebraically,
\begin{align}
\langle O_{\rm SC}(x)O_{\rm SC}(0)\rangle
&\sim x^{-2/K},\nonumber\\
\langle \rho(x)\rho(0)\rangle_{2k_F}
&\sim \cos\left[\left(\frac{\pi}{a_0}+\operatorname{sgn}(\mu)\frac{\delta q}{2}\right)x\right]x^{-2K},
\end{align}
where the $2k_F$ part of the density is modulated at $2k_F=\pi/a_0+\operatorname{sgn}(\mu)\,\pi n_s$, so that its deviation from commensurability is $\delta q/2$. These correlations are consistent with the absence of long-range order and the underlying $c=1$ conformal field theory.

\begin{figure}
    \centering
    \includegraphics[width=\linewidth]{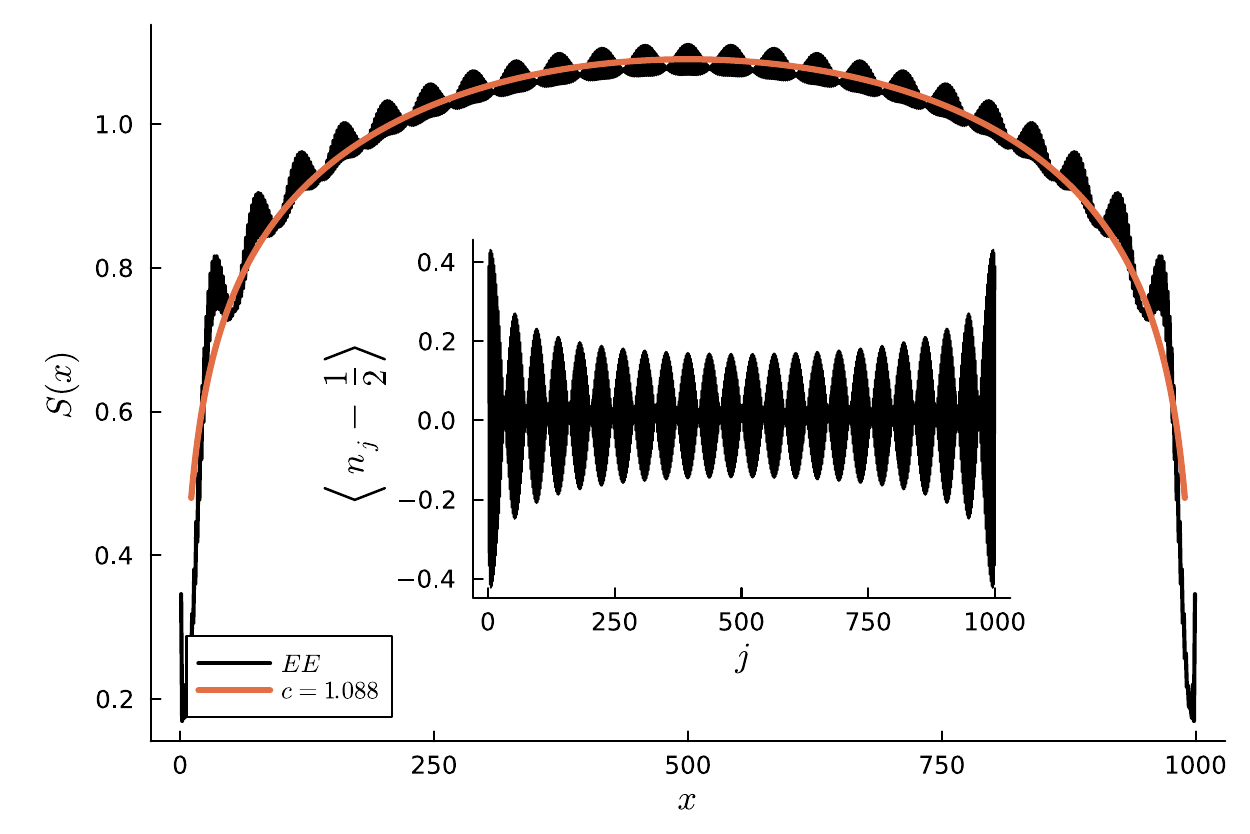}
    \caption{Entanglement entropy $S(x)$ in the incommensurate Luttinger-liquid phase for $\mu = 2.5$, $U = 1.5$, $t = 1$, and $\Delta = 0.5$ on an open chain with $N = 1000$ sites. The DMRG data are well described by the Calabrese-Cardy conformal scaling form, yielding a fitted central charge $c \simeq 1.088$, consistent with a gapless $c=1$ Luttinger liquid. To reduce boundary-induced oscillatory corrections, the fit excludes the outer $1\%$ of bonds near each edge of the chain. \textit{Inset}: local density profile $\langle n_j - 1/2 \rangle$ showing weak incommensurate density oscillations characteristic of the incommensurate phase.}
    \label{fig:EE-IC}
\end{figure}

To confirm the gapless nature of this regime, we compute the bipartite entanglement entropy using DMRG with open boundary conditions. Conformal invariance predicts the Calabrese-Cardy scaling form~\cite{calabrese2004entanglement}
\begin{equation}
S(x) = \frac{c}{6} \ln\left[ \frac{L}{\pi}
\sin\left( \frac{\pi x}{L} \right) \right] +s_0.
\label{eq:CC_scaling}
\end{equation}
Fitting the numerical entanglement entropy to Eq.~\eqref{eq:CC_scaling} yields a central charge consistent with $c=1$, confirming that the intermediate phase is a gapless incommensurate Luttinger liquid. Representative DMRG results for $\mu=2.5$ and $U=1.5$ on an open chain with $N=1000$ sites are shown in Fig.~\ref{fig:EE-IC}, where the entanglement entropy closely follows the conformal scaling prediction with fitted central charge $c\simeq1.088$, consistent with the expected $c=1$ theory up to finite-size and incommensurate Friedel oscillations. The inset density profile $\langle n_j-1/2\rangle$ exhibits weak oscillations with a continuously varying incommensurate wavevector, associated with a finite soliton density and the absence of commensurate charge locking.

Approaching the Pokrovsky--Talapov transition from the incommensurate side, the Luttinger parameter of the emergent soliton liquid approaches the universal value $K_{\mathrm{PT}}=\frac14$~\cite{giamarchi1997mott,giamarchi2003quantum}. This follows from the universal dilute-soliton description of the commensurate-incommensurate transition. Near threshold, the soliton gas becomes asymptotically equivalent to a free spinless Fermi gas, whose bosonized Luttinger parameter is $K_s=1$. Since each elementary soliton carries the winding
$|\Delta\phi|=\frac{\pi}{2}$,
the canonically normalized soliton field is related to the original bosonization field by $\varphi_s=2\phi$, which rescales the Gaussian stiffness and yields the universal Pokrovsky--Talapov value $K_{\mathrm{PT}}=1/4$ in the original normalization of Eq.~\eqref{eq:LL_hamiltonian}. This should be contrasted with the Berezinskii--Kosterlitz--Thouless transition
bounding the incommensurate phase on the superconducting side, which occurs when the pairing perturbation $\cos(2\theta)$ becomes marginal at $K_{\mathrm{BKT}}=\frac12$~\cite{giamarchi2003quantum,giamarchi1997mott}.

\begin{figure}
\centering

\begin{tikzpicture}

\begin{axis}[width=9cm,height=9cm,xmin=0,
    xmax=2.35, ymin=0, ymax=4.0, xlabel={$U/t$},
    ylabel={$\mu/t$},axis lines*=box,
    xtick={0,0.5,1,1.5,2},ytick={0,1,2,3,4},
    tick style={black, thick}, every axis plot/.append style={line width=1.2pt,
        line cap=round,line join=round
    },clip=true,
    ]
\addplot[name path=leftaxis, draw=none] coordinates {
    (0,0)
    (0,4.0)
};

\addplot[name path=rightaxis, draw=none] coordinates {
    (2.35,0)
    (2.35,4.0)
};

% TSC--IC boundary
\addplot[name path=tSCIC, draw=none] coordinates {
    (0.75,0.0)
    (0.755,0.5)
    (0.78,1.0)
    (0.85,1.5)
    (0.89,2.0)
    (0.93,2.5)
    (1.00,3.0)
    (1.10,3.5)
    (1.17,4.0)
};

% IC--CDW boundary
\addplot[name path=ICCDW, draw=none] coordinates {
    (0.75,0.0)
    (0.995,0.5)
    (1.186,1.0)
    (1.365,1.5)
    (1.54,2.0)
    (1.71,2.5)
    (1.88,3.0)
    (2.00,3.5)
    (2.20,4.0)
};

% TI boundary
\addplot[name path=TI, draw=none] coordinates {
    (0.00,2.0)
    (0.01,2.0)
    (0.14,2.5)
    (0.27,3.0)
    (0.39,3.5)
    (0.52,4.0)
};

\addplot[name path=TIleft, draw=none] coordinates {
    (0.00,2.0)
    (0.00,4.0)
};

% TSC
\addplot[fill=blue!18,draw=none]
fill between[of=leftaxis and tSCIC];

% IC
\addplot[fill=green!18,draw=none]
fill between[of=tSCIC and ICCDW];

% CDW
\addplot[fill=red!16, draw=none]
fill between[of=ICCDW and rightaxis];

% TI
\addplot[fill=yellow!22,draw=none]
fill between[of=TIleft and TI];

% TSC--IC boundary
\addplot[thick,blue,smooth,mark=*,mark size=2.2pt,
    mark options={fill=blue, draw=blue    }
] coordinates {
    (0.75,0.0)
    (0.755,0.5)
    (0.78,1.0)
    (0.85,1.5)
    (0.89,2.0)
    (0.93,2.5)
    (1.00,3.0)
    (1.10,3.5)
    (1.17,4.0)
};

% IC--CDW boundary
\addplot[thick, blue, densely dotted,
    smooth, mark=*, mark size=2.2pt,
    mark options={fill=white, draw=blue}
] coordinates {
    (0.75,0.0)
    (0.995,0.5)
    (1.186,1.0)
    (1.365,1.5)
    (1.54,2.0)
    (1.71,2.5)
    (1.88,3.0)
    (2.00,3.5)
    (2.20,4.0)
};

% TI boundary
\addplot[thick,blue,densely dotted,
    smooth,mark=*,  mark size=2.2pt,
    mark options={ fill=white,draw=blue}
] coordinates {
    (0.00,2.0)
    (0.01,2.0)
    (0.14,2.5)
    (0.27,3.0)
    (0.39,3.5)
    (0.52,4.0)
};

\node[font=\Large] at (axis cs:0.38,1.65) {TSC};

\node[font=\Large] at (axis cs:1.1,1.90) {IC};

\node[font=\Large] at (axis cs:1.95,1.25) {CDW};

\node[font=\Large] at (axis cs:0.18,3.35) {TI};

\path
(axis cs:0.97,2.45)
node[rotate=83, blue,anchor=south,font=\small] {BKT $K=\frac{1}{2}$};

\path
(axis cs:1.75,2.50)
node[
    rotate=60,
    blue,
    anchor=south,
    font=\small
] {PT $|\mu|/2=M_s$};

\path
(axis cs:0.4,2.90)
node[ rotate=66, blue, anchor=south,  font=\small
] {Ising};

\addplot[ only marks,mark=star, mark size=4pt,
    yellow!80!black, draw=black] coordinates {
    (0.75,0.0)};

\end{axis}

\end{tikzpicture}
\caption{Phase diagram showing the topological superconducting (TSC), trivial insulating (TI), incommensurate Luttinger-liquid (IC), and commensurate charge-density-wave (CDW) phases. The overall structure of the phase diagram is motivated by the weak-coupling bosonization analysis discussed above, while the symbols show phase-transition points obtained from DMRG calculations at $t=1$ and $\Delta=0.5$. The lines connecting the DMRG data points serve as guides to the eye. The solid line denotes the IC--TSC BKT boundary for $\mu>0$ and terminates at the particle-hole-symmetric eight-vertex point, while the dotted lines denote the Pokrovsky--Talapov and Ising boundaries. The star marks the particle-hole-symmetric point, whose position $U_c=\frac{t+\Delta}{2}=0.75$ is known exactly from Eq.~\eqref{eq:exact_Uc} and is reproduced by the DMRG calculations. The DMRG calculations were performed on open chains with $N=1000$ sites using up to $35$ sweeps, a maximum bond dimension $\chi=1600$, a truncation cutoff of $10^{-12}$, and an energy convergence criterion $\Delta E<10^{-12}$.}
\label{fig:phase-diagram}
\end{figure}
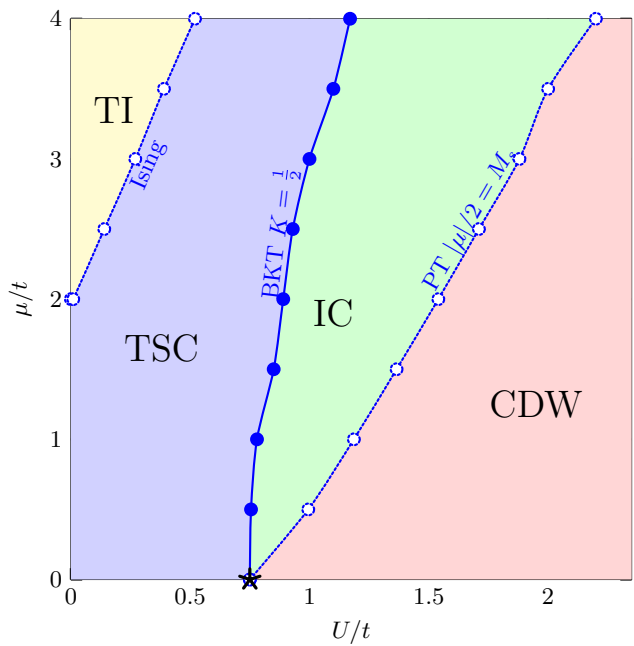

\subsection{Ising transition to the trivial phase}
\label{sec:trivial}

A second mechanism for destroying the topological superconducting phase is the increase of the chemical potential. Physically, this drives the system into a trivial insulating regime in which the characteristic topological boundary excitations disappear. In the noninteracting limit, the bulk quasiparticle gap closes at $|\mu|=2t$, separating the topological superconducting phase from a trivial band insulator.  
In the interacting system, this critical point is generally renormalized away from its noninteracting value, although weak interactions preserve the same underlying gap-closing mechanism so long as the transition remains continuous.

Unlike the interaction-driven transitions discussed above, this critical point does not arise from competing cosine perturbations, and it lies outside the regime of validity of the expansion about $k_F^0=\pi/(2a_0)$. Instead, it is directly characterized in the fermionic language. In the noninteracting limit, the Bogoliubov  quasiparticle spectrum is
\begin{equation}
E_k=\sqrt{\left(2t\cos k+\mu\right)^2+4\Delta^2\sin^2 k},
\end{equation}
which vanishes at $k=\pi$ when $\mu=2t$ (and at $k=0$ when $\mu=-2t$). Setting $k=\pi+q$ gives
\begin{equation} 
2t\cos k+\mu\simeq \mu-2t+tq^2, \quad 2\Delta\sin k\simeq-2\Delta q,
\end{equation}
and
\begin{equation}
E(q)\simeq 2|\Delta|\,|q|
\end{equation}
at the critical point $\mu=2t$ for $\Delta\neq0$. The pairing thus keeps the critical dispersion linear. The resulting critical theory contains a single massless Majorana mode, has dynamical exponent $z=1$, and central charge
$c=1/2$, and belongs to the Ising universality class. Weak interactions shift the location of the transition and renormalize the velocity, but do not change this universality class, provided that no additional gapless mode or multicritical point intervenes, consistent with the Ising transition lines identified in Ref.~\cite{chepiga2023eight}.

In the number-conserving limit $\Delta=0$, the pairing term is absent and the same band-edge expansion gives $E(q)\sim q^2$, corresponding to a dilute-fermion transition with $z=2$. This limit, however, does not describe a transition out of a topological superconducting phase. Beyond the transition at nonzero pairing, the system is a trivial gapped phase adiabatically connected to the empty or fully filled band.

\begin{figure}[H]
    \centering    \includegraphics[width=\linewidth]{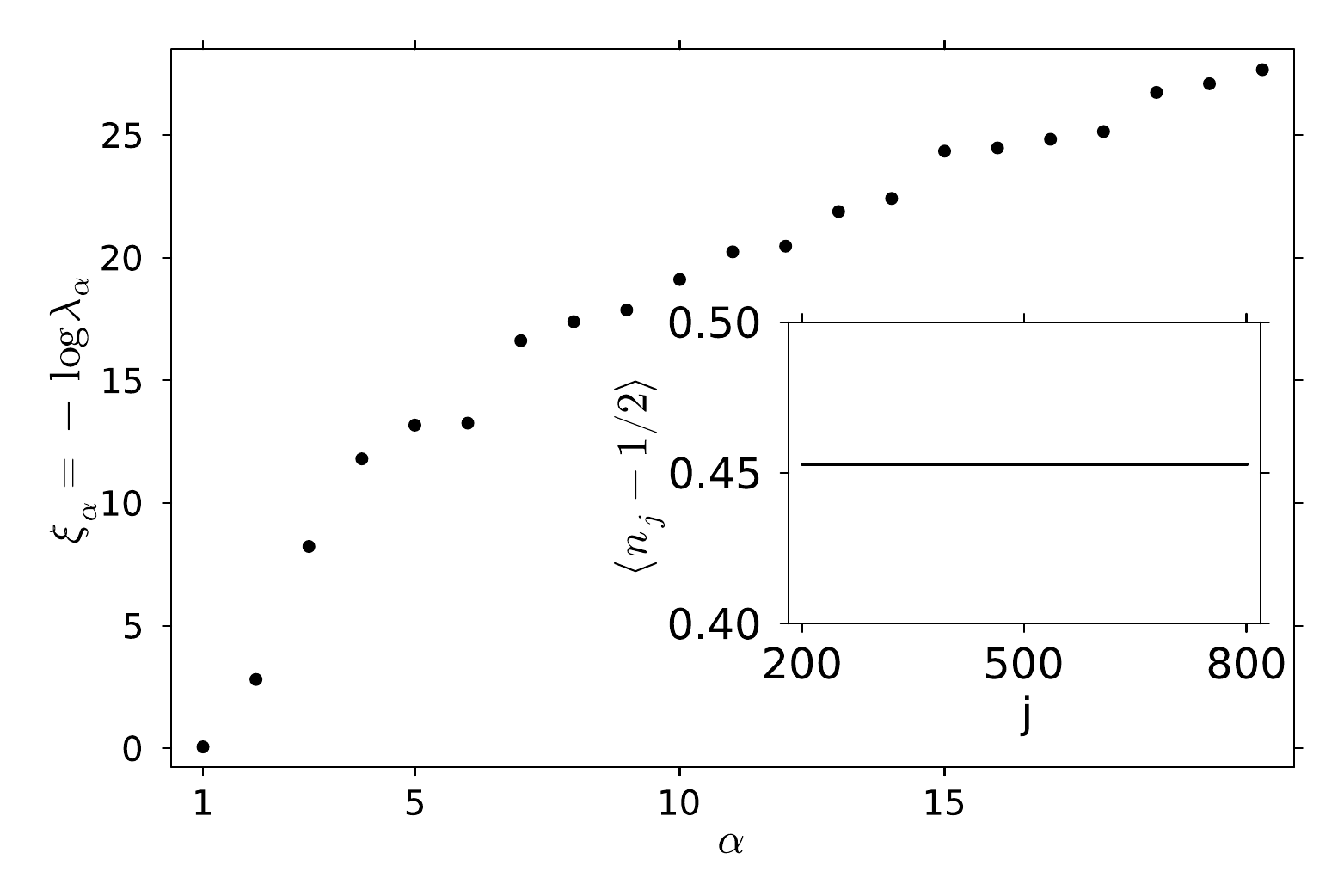}
    \caption{Representative DMRG results in the trivial insulating regime for $\mu = 4$, $U = 0.5$, $t = 1$, and $\Delta = 0.5$ on an open chain with $N = 1000$ sites. \textit{Main panel}: entanglement spectrum $\xi_\alpha = -\log \lambda_\alpha$, showing the absence of the near-double degeneracy characteristic of the topological superconducting phase. \textit{Inset}: local density profile $\langle n_j - 1/2 \rangle$, which remains nearly uniform throughout the bulk and approaches the fully filled limit due to the large chemical potential.}
    \label{fig:trivial-phase}
\end{figure}

The DMRG calculations identify the trivial insulating regime through several complementary signatures. The local density profile $\langle n_j-1/2\rangle$ becomes nearly uniform throughout the bulk and approaches the fully filled limit set by the large chemical potential, indicating the absence of both commensurate charge ordering and incommensurate density modulations. Simultaneously, the entanglement spectrum becomes fully nondegenerate, consistent with the disappearance of topological boundary excitations and the absence of protected edge modes. In Fig.~\ref{fig:trivial-phase}, we show DMRG results for the trivial insulating phase at $\mu=4$ and $U=0.5$ on an open chain with $N=1000$ sites.

\subsection{Summary of the phase diagram}

Combining the bosonization analysis with DMRG calculations at fixed $t=1$ and $\Delta=0.5$ yields the schematic $(U,\mu)$ phase diagram shown in Fig.~\ref{fig:phase-diagram}. The phase diagram contains four phases. These are the topological superconducting (TSC) and commensurate charge-density-wave Mott (CDW) phases, together with an incommensurate Luttinger liquid (IC) and a trivial insulator (TI). The CDW--IC and IC--TSC boundaries are governed by Pokrovsky--Talapov and Berezinskii--Kosterlitz--Thouless criticality, respectively, whereas the TSC--TI boundary is in the  Ising universality class.
At $\mu=0$, the topological superconducting and charge-ordered phases meet directly at the multicritical point Eq.~\eqref{eq:exact_Uc}, with the eight-vertex exponents of Eq.~\eqref{eq:eightvertex_exponents}. Away from half filling, the direct transition is split by an intervening incommensurate Luttinger liquid, so that this point is the endpoint at which the Pokrovsky--Talapov and
Kosterlitz--Thouless boundaries merge, producing the structure of Fig.~\ref{fig:phase-diagram}.

\section{Field Theory of the Topological Phase}\label{sec:field-theory-topological}

Sections~\ref{sec:RG} and~\ref{sec:Phase_transtions} addressed the bulk problem, namely the competition between the pairing perturbation $\cos(2\theta)$, Umklapp locking, and doping, and the critical lines separating the resulting phases. We now turn to the topological superconducting phase itself. The bulk analysis done above shows that the pairing perturbation flows to strong coupling, where it pins $\theta$ and opens a spectral gap. The existence of this gap does not by itself establish the topological character of the phase, since the trivial insulator and the commensurate charge-density-wave phase are also gapped. The topological superconducting phase is instead distinguished by its boundary excitations and by the fermion-parity structure of the many-body Hilbert space~\cite{fidkowski2010effects,fidkowski2011topological}. These properties are not determined by the bulk renormalization-group flow and require an explicit treatment of the open-chain theory. We study the boundary excitations using the band-edge Ising theory and finite-size bosonization, retaining the compact zero modes needed to describe the fermion-parity sectors and their finite-size splitting. We then construct a dual fermionic representation in which the same gapped bulk sector has a conventional symmetry-breaking description with a local order parameter. This provides a continuum realization of the SPT--SSB correspondence~\cite{tang2026topological}.

A defining property of the topological superconducting phase is the presence of exponentially localized boundary excitations. In the bosonized description, the bulk superconducting gap originates from the
relevant pairing perturbation $\cos(2\theta)$, which flows to strong coupling and pins the superconducting phase field $\theta$. Unlike the charge-density-wave phase, where domain walls are directly represented by kinks of the field $\phi$, the corresponding boundary structure in the
superconducting phase is encoded in the dual field $\theta$ and is less transparent.

Because the pairing perturbation breaks the microscopic $U(1)$ symmetry down to fermion parity, the compactification should be specified for the pair $(\phi,\theta)$ rather than for $\theta$ alone. Single-valuedness of the fermion operators in Eq.~\eqref{eq:bosonization_dictionary} and of the density vertex $e^{2i\phi}$ requires the identification lattice
\begin{equation}
(\phi,\theta)\sim(\phi+\pi,\theta+\pi),\quad
(\phi,\theta)\sim(\phi,\theta+2\pi).
\label{eq:identification_lattice}
\end{equation}
The shift $\theta\rightarrow\theta+\pi$ by itself is not an identification. It leaves the pairing term $\cos(2\theta)$ unchanged but sends $\psi_r\rightarrow-\psi_r$, and hence acts as fermion parity. This distinction between compactification and fermion parity becomes important for an open chain and will be made explicit below when we construct the finite-size Hilbert space and its parity sectors.

A more transparent description is obtained after refermionization. Using the bosonization dictionary
Eq.~\eqref{eq:bosonization_dictionary}, the pairing interaction
\begin{equation}
H_\Delta = \frac{v y_\Delta}{2\pi a_0^2} \int \mathrm{d}x\,\cos(2\theta)
\label{eq:pairing_bosonized_edge}
\end{equation}
maps onto the $p$-wave pairing operator
\begin{equation}
\cos(2\theta)  \longleftrightarrow \psi_R\psi_L+\psi_L^\dagger\psi_R^\dagger,
\label{eq:fermionization_pairing}
\end{equation}
up to a phase convention for the fermionic fields.

Introducing the Nambu spinor
\begin{equation}
\Psi(x)=
\begin{pmatrix}
\psi_R(x)\\
\psi_L^\dagger(x)
\end{pmatrix},
\label{eq:Nambu_spinor}
\end{equation}
the low-energy Hamiltonian takes the Dirac form
\begin{equation}
H=\int\mathrm{d}x \Psi^\dagger(x) \left[ -i v\sigma^z\partial_x+m_\Delta(x)\sigma^x \right] \Psi(x),
\label{eq:Dirac_H_edge}
\end{equation}
where $m_\Delta(x)$ denotes the effective continuum pairing-mass profile.  This reduced block is useful for describing spectral matching, but its mass kink alone does not determine the number of physical Majorana modes. The full BdG problem, together with the boundary conditions, determines the topological character, not the sign of the uniform pairing coupling $y_\Delta$ alone.

\subsection{Band-edge Ising description of the topological edge mode}
\label{sec:boundary_physics}

The $k_F$-linearized pairing theory is useful for describing the superconducting gap and the decay scale of boundary excitations. However, a sign change of the pairing mass $m_\Delta$ does not by itself distinguish the topological and trivial phases of the Kitaev chain. The physical single-Majorana boundary mode is instead most transparently obtained from the band-edge theory near the topological transition, where the relevant mass changes sign between the topological and trivial regions.

The band-edge theory discussed in Sec.~\ref{sec:trivial} makes this explicit. Expanding the
bulk spectrum about $k=\pi$ gives
\begin{equation}
2t\cos(\pi+q)+\mu
\simeq \mu-2t+tq^2,
\quad
2\Delta\sin(\pi+q)\simeq-2\Delta q.
\end{equation}
The resulting low-energy theory at the transition is a single Ising
Majorana theory with velocity
\begin{equation}
v_{\rm I}=2|\Delta|a_0
\end{equation}
and mass
\begin{equation}
m_{\rm I}(x)= \mu(x)-2t,
\label{eq:Ising_mass_profile}
\end{equation}
so that $E^2(q)=m_{\rm I}^2+v_{\rm I}^2q^2/a_0^2$. The term $tq^2$ can be neglected provided $|\mu-2t|\ll4\Delta^2/t$, which sets the regime of validity of this description.  The analogous expansion at
$k=0$ describes the transition at $\mu=-2t$.

In a real Majorana basis with $\{\xi_a(x),\xi_b(y)\}=\delta_{ab}\delta(x-y)$, the band-edge theory can be written as
\begin{equation}
H_{\rm I}=\frac{i}{2}\int\mathrm{d}x\left[
v_{\rm I}\xi_R\partial_x\xi_R
-v_{\rm I}\xi_L\partial_x\xi_L
+2m_{\rm I}(x)\xi_R\xi_L\right].
\label{eq:Ising_Majorana_H}
\end{equation}
A spatial kink of $m_{\rm I}(x)$ through zero binds one
self-conjugate zero mode~\cite{jackiw1976solitons}.  For a sharp kink,
\begin{equation}
m_{\rm I}(x)=M_{\rm I}\operatorname{sgn}(x),\quad M_{\rm I}>0,
\end{equation}
where the topological region ($m_{\rm I}<0$) lies at $x<0$, so that the kink describes the right end of a topological region. Denoting the spatial envelope of the Majorana zero mode by $f_{\rm I}(x)$, one finds
\begin{equation}
f_{\rm I}(x)\propto
\exp\left[-\frac{1}{v_{\rm I}}\int_0^x
\mathrm{d}x'\,m_{\rm I}(x')\right]
=e^{-M_{\rm I}|x|/v_{\rm I}},
\quad
\xi_{\rm I}=\frac{v_{\rm I}}{M_{\rm I}}.
\label{eq:Ising_edge_profile}
\end{equation}
The corresponding boundary operator is
\begin{equation}
\gamma_{\rm I}=\int\mathrm{d}x\,f_{\rm I}(x)
\left[\xi_R(x)\pm\xi_L(x)\right],
\quad
\gamma_{\rm I}^\dagger=\gamma_{\rm I},
\quad
\gamma_{\rm I}^2=1,
\label{eq:Ising_edge_operator}
\end{equation}
where the relative sign is fixed by the orientation of the mass kink. The upper sign applies to the kink above, while the lower sign, with the same envelope, applies to the opposite orientation, as at the left end of the chain. The condition $\gamma_{\rm I}^2=1$ fixes the normalization $\int\mathrm{d}x\,f_{\rm I}^2=1$. As a check, $\xi_{\rm I}=2|\Delta|a_0/(2t-\mu)$ agrees with the $\mu\to2t^-$ limit of the exact lattice decay length obtained from the slower root in Eq.~\eqref{eq:majorana_roots} below.

This construction gives one Majorana mode at an interface between the topological and trivial phases. For the physical open Kitaev chain, the detailed spatial profile of the boundary mode is obtained from the unfolded continuum theory in  Sec.~\ref{sec:finite_size_bosonization}. The band-edge description also makes clear why the topological--trivial transition is governed by a single massless Majorana mode with $z=1$ and $c=1/2$.

Interactions renormalize the bulk mass scale.  From the weak-coupling RG
flow of the pairing coupling, Eq.~\eqref{eq:RG_pairing}, the superconducting
correlation length scales as
\begin{align}
M(K)& \sim D\,|y_{\Delta,0}|^{1/(2-1/K)},\\
\xi_{\rm edge}&\sim\frac{v}{M(K)}
\sim a_0\,|y_{\Delta,0}|^{-1/(2-1/K)}.
\label{eq:M_scaling}
\end{align}
At $K=1$ this gives the weak-pairing continuum result.  As the
Berezinskii--Kosterlitz--Thouless transition $K\rightarrow1/2^+$ is
approached, the fixed-$K$ power law is replaced by the BKT essential
singularity in Eq.~\eqref{eq:BKT_gap}, and the correlation length diverges.
\subsection{Finite-size bosonization and the boundary Hilbert space}
\label{sec:finite_size_bosonization}

The semiclassical analysis above identifies the localized boundary excitation and relates its decay length to the superconducting bulk gap.  On a finite open chain, bosonization additionally resolves the structure of the many-body Hilbert space where the boundary conditions select the allowed field modes, a compact zero-mode pair remains, and fermion parity separates the Hilbert space into two sectors.  We first establish this structure for arbitrary $K$.  We then specialize to $K=1$, where the finite-size boundary problem can be solved exactly within the continuum theory, generalizing the construction of Ref.~\cite{chua2020majorana} to finite chemical potential.  Finally, we return to $K\neq1$ and describe the interacting boundary sector directly within the sine-Gordon theory.

\subsubsection{Open-boundary quantization}

For a hard wall, the chiral fields obey~\cite{fabrizio1995interacting}
\begin{align}
\psi_L(0)&=-\psi_R(0),\nonumber\\
\psi_L(L)&=-e^{2ik_F^0L}\psi_R(L).
\label{eq:open_chiral_bc}
\end{align}
For an open chain of $N$ sites, the hard walls sit at the fictitious sites $j=0$ and $j=N+1$, so that $L=(N+1)a_0$.
With the bosonization convention of Eq.~\eqref{eq:bosonization_dictionary}, this fixes the density field at the two ends,
\begin{align}
\phi(0) &=\frac{\pi}{2}\pmod{\pi}, \nonumber\\
\phi(L) &=\frac{\pi}{2}+k_F^0L\pmod{\pi}.
\label{eq:phi_bc_finite}
\end{align}
The modulo-$\pi$ condition follows from the hard-wall relation, which fixes $e^{2i\phi}$ at the boundary.  The bosonic field itself remains compact with period $2\pi$, and the different allowed windings between the two endpoints are encoded by the global charge sector.

For a hard-wall boundary, the dual condition is the kinematic Neumann condition associated with the compact Dirichlet condition on $\phi$.  Since
\begin{equation}
\partial_t\phi=vK \partial_x\theta, \quad j=\frac{1}{\pi}\partial_t\phi,
\end{equation}
one obtains
\begin{equation}
\partial_x\theta(0) = \partial_x\theta(L) = 0.
\label{eq:theta_bc_finite}
\end{equation}
Thus $\phi$ obeys compact Dirichlet boundary conditions and $\theta$ obeys Neumann boundary conditions.

Choose a $c$-number background $\phi_{\rm b}(x)$ implementing one
representative of Eq.~\eqref{eq:phi_bc_finite}.  A finite-size expansion consistent with the open boundaries is
\begin{align}
\theta_a(x) &= \theta_0 + \sum_{n=1}^{\infty} \frac{e^{-n\pi a/(2L)}}{\sqrt{Kn}} \left(a_n+a_n^\dagger\right) \cos\left(\frac{n\pi x}{L}\right),
\label{eq:theta_open_modes}\\
\phi_a(x) &= \phi_{\rm b}(x) -\frac{\pi Q}{L}x \nonumber\\
&\quad +i\sum_{n=1}^{\infty} e^{-n\pi a/(2L)} \sqrt{\frac{K}{n}}
\left(a_n^\dagger-a_n\right) \sin\left(\frac{n\pi x}{L}\right),
\label{eq:phi_open_modes}
\end{align}
where
\begin{equation}
[a_n,a_m^\dagger]=\delta_{nm}, \quad [\theta_0,Q]=-i, \quad Q\in\mathbb Z.
\label{eq:finite_mode_algebra}
\end{equation}
The sine modes vanish at the boundaries, while the derivatives of the cosine modes vanish there.  The term $-\pi Qx/L$ changes $\phi(L)$ only by an integer multiple of $\pi$ and labels the allowed compact winding sectors. This regularized open-boundary expansion has the same structure as the one used for the interacting SSH chain in Ref.~\cite{jin2023bosonization}.

The mode expansion also makes the canonical algebra explicit.  One finds
\begin{align}
[\phi_a(x),\partial_y\theta_a(y)] &= i\pi \delta_{D,a}(x,y),
\label{eq:regulated_open_commutator} \\
\delta_{D,a}(x,y) &= \frac{2}{L} \sum_{n=1}^{\infty} e^{-n\pi a/L}
\nonumber\\
&\quad\times \sin\left(\frac{n\pi x}{L}\right) \sin\left(\frac{n\pi y}{L}\right).
\label{eq:dirichlet_kernel}
\end{align}
As $a\rightarrow0^+$, $\delta_{D,a}$ becomes the identity kernel on the Dirichlet fluctuation space.  In the interior,
\begin{equation}
\delta_D(x,y)=\delta(x-y), \quad 0<x,y<L,
\end{equation}
and the bulk canonical relation
\begin{equation}
[\phi(x),\partial_y\theta(y)] = i\pi\delta(x-y)
\end{equation}
is recovered.  At the endpoints, the Dirichlet kernel vanishes because the boundary value of $\phi$ is fixed and is not an independent dynamical coordinate.  The Dirichlet and Neumann boundary conditions are fully compatible with the canonical algebra.

The zero-momentum part of the theory is the conjugate pair $(\theta_0,Q)$. Accordingly,
\begin{equation}
\mathcal H_{\rm open} = \mathcal H_{\rm zero}\otimes\mathcal H_{\rm osc}.
\label{eq:open_hilbert_space}
\end{equation}
This factorization is kinematic; the pairing perturbation couples the zero
mode to the oscillator sector.  For
\begin{equation}
Q|q\rangle=q|q\rangle, \quad q\in\mathbb Z,
\end{equation}
the compact phase operator translates between neighboring charge sectors,
\begin{align}
e^{i\theta_0}|q\rangle&=|q-1\rangle,\nonumber\\
e^{-i\theta_0}|q\rangle&=|q+1\rangle.
\label{eq:zero_mode_translation}
\end{align}

The zero-mode energy follows from the spatial integral of $\partial_x\phi$.  For the background used in Eq.~\eqref{eq:phi_open_modes}, define its average gradient by
\begin{equation}
g=\frac{\phi_{\rm b}(L)-\phi_{\rm b}(0)}{L}.
\end{equation}
The mode expansion then gives
\begin{equation}
\int_0^L\mathrm{d}x\,\partial_x\phi = gL-\pi Q .
\end{equation}
Using the chemical-potential coupling introduced above,
\begin{equation}
H_\mu= \frac{\mu}{\pi} \int_0^L\mathrm{d}x\,\partial_x\phi ,
\end{equation}
the Gaussian energy of the charge sector becomes
\begin{equation}
H_{0,\rm zero} = \frac{\pi v}{2KL}(Q-\bar Q)^2+\mathrm{const.},
\label{eq:zero_mode_charging}
\end{equation}
where
\begin{equation}
\bar Q= \frac{L}{\pi} \left( g+\frac{K\mu}{v} \right).
\end{equation}
Thus the chemical potential shifts the relative energies of the integer $Q$ sectors, whereas the oscillator modes and their canonical algebra are unchanged.
Superconducting pairing changes $Q$ by two units and preserves the fermion parity
\begin{equation}
P=e^{i\pi Q},
\label{eq:parity_zero_sector}
\end{equation}
where the possible background phase associated with $gL/\pi$ has been
absorbed into the definition of the parity convention.
When comparing with the lattice, this convention is fixed by identifying $Q$ with the total fermion number $F$. The total fermion number differs from $\rho_0L+Q_\phi$ by the boundary charge carried by the $2k_F$ part of the density, Eq.~\eqref{eq:density_bosonized}, near the two hard walls. Requiring that, at $\Delta=0$, $K=1$, and $\mu=0$, Eq.~\eqref{eq:zero_mode_charging} reproduces the free open chain, whose ground state is unique for even $N$ and doubly degenerate for odd $N$, fixes this boundary charge to $-1/2$, so that the charging energy is centered at $F=N/2+KL\mu/(\pi v)$ rather than at $k_F^0L/\pi+KL\mu/(\pi v)$.
The Hilbert space separates into the two invariant sectors
\begin{equation}
\mathcal H_{\rm open} = \mathcal H_+\oplus\mathcal H_-,
\label{eq:parity_hilbert_decomposition}
\end{equation}
with $P=+1$ in $\mathcal H_+$ and $P=-1$ in $\mathcal H_-$.  A charge-one vertex exchanges these sectors,
\begin{equation}
P e^{\pm i\theta_0}P^{-1} = -e^{\pm i\theta_0},
\label{eq:parity_vertex}
\end{equation}
or equivalently
\begin{equation}
P\theta_0P^{-1} = \theta_0+\pi\pmod{2\pi}.
\label{eq:parity_zero_mode}
\end{equation}

\subsubsection{Finite-size boundary modes at $K=1$}

At $K=1$ the boundary problem can be solved exactly within the continuum theory. Since the fermions are free, it is convenient to linearize around the actual Fermi momentum $k_F$ at chemical potential $\mu$,
\begin{equation}
2t\cos(k_Fa_0)+\mu=0,
\quad
v=2ta_0\sin(k_Fa_0),
\label{eq:finite_mu_kF_vF}
\end{equation}
so that the pairing term gives the mass
\begin{equation}
M_\Delta=2\Delta\sin(k_Fa_0),
\label{eq:finite_mu_pairing_mass}
\end{equation}
which we take positive. The continuum Hamiltonian is then
\begin{align}
H_{K=1}&=\int_0^L\mathrm{d}x\Big[-iv\left(\psi_R^\dagger\partial_x\psi_R-\psi_L^\dagger\partial_x\psi_L\right)\nonumber\\
&\quad+M_\Delta\left(\psi_R\psi_L+\psi_L^\dagger\psi_R^\dagger\right)\Big],
\label{eq:continuum_K1_H}
\end{align}
with the hard-wall conditions of Eq.~\eqref{eq:open_chiral_bc}, where $k_F^0$ is replaced by $k_F$.

We use the standard unfolding trick. The condition at $x=0$ allows us to write the left mover as a right mover on the negative axis,
\begin{equation}
\psi_L(x)=-\psi_R(-x),
\end{equation}
so that the whole problem is described by a single right mover $\psi_R(y)$ with $-L<y<L$. The condition at $x=L$ then becomes a twisted periodicity,
\begin{equation}
\psi_R(y+2L)=e^{-2ik_FL}\,\psi_R(y).
\label{eq:twisted_bc}
\end{equation}
We look for Bogoliubov quasiparticles of the form
\begin{equation}
\Gamma=\int_{-L}^{L}\mathrm{d}y\left[\alpha(y)\,\psi_R^\dagger(y)+\beta(y)\,\psi_R(y)\right].
\end{equation}
Imposing $[H_{K=1},\Gamma]=E\,\Gamma$ and writing $\gamma(y)=\beta(-y)$, the two amplitudes obey a single Dirac equation,
\begin{equation}
E\begin{pmatrix}\alpha\\ \gamma\end{pmatrix}
=\left[-iv\sigma^z\partial_y+M_\Delta\,{\rm sgn}(y)\,\sigma^x\right]\begin{pmatrix}\alpha\\ \gamma\end{pmatrix},
\label{eq:unfolded_dirac}
\end{equation}
on a ring of length $2L$ with the twisted condition $(\alpha,\gamma)(y+2L)=e^{-2ik_FL}(\alpha,\gamma)(y)$ inherited from Eq.~\eqref{eq:twisted_bc}. Notice that in the unfolded picture, each boundary of the chain becomes a mass domain wall, located at $y=0$ for the left edge and at $y=\pm L$ for the right edge.

A single sign change of the mass binds exactly one zero-energy solution. For the left edge, the normalizable solution of Eq.~\eqref{eq:unfolded_dirac} with $E=0$ is
\begin{equation}
\alpha(y)=a\,e^{-|y|/\xi},\quad \gamma(y)=-ia\,e^{-|y|/\xi},\quad \xi=\frac{v}{M_\Delta},
\label{eq:continuum_coherence_length}
\end{equation}
and requiring $\Gamma^\dagger=\Gamma$ fixes the phase $a=e^{i\pi/4}/\sqrt{\xi}$. This gives a single Majorana operator
\begin{equation}
\Gamma_L=\frac{1}{\sqrt{\xi}}\int\mathrm{d}y\,e^{-|y|/\xi}\left[e^{i\pi/4}\psi_R^\dagger(y)+e^{-i\pi/4}\psi_R(y)\right],
\label{eq:continuum_majorana_L}
\end{equation}
which satisfies $\Gamma_L^\dagger=\Gamma_L$ and $\Gamma_L^2=1$. Changing $a$ by a complex factor only multiplies $\Gamma_L$ by the same factor, so each edge hosts exactly one Majorana mode. This does not contradict the discussion of Sec.~\ref{sec:boundary_physics}. A sign change of the pairing mass inside the wire involves two independent fields, $\psi_R$ and $\psi_L$, and binds two Majorana modes, whereas at a physical edge the hard wall ties $\psi_L$ to $\psi_R$, leaving a single field and hence a single Majorana mode. The right-edge operator $\Gamma_R$ is constructed in the same way around $y=L$, with its phase fixed by the twist in Eq.~\eqref{eq:twisted_bc}. The open chain hosts two Majorana modes in total, one at each end, as in the Kitaev chain.

Folding back, the amplitude of $c^\dagger(x)$ in $\Gamma_L$ is proportional to $e^{ik_Fx}\alpha(x)-e^{-ik_Fx}\alpha(-x)$. The same applies at the right edge, so that
\begin{align}
\Psi_L(x)&\sim e^{-x/\xi}\sin(k_Fx),\nonumber\\
\Psi_R(x)&\sim e^{-(L-x)/\xi}\sin\left[k_F(L-x)\right],
\label{eq:continuum_mu_edge_profiles}
\end{align}
so that the edge modes are exponentially localized standing waves. Notice that for the nearest-neighbor model the decay length $\xi=v/M_\Delta=ta_0/\Delta$ does not depend on $k_F$ within this linearized theory.

For a finite chain, the Majorana modes localized near the two ends have exponentially decaying wavefunctions that overlap through the bulk. This overlap hybridizes the two edge modes and produces a small finite-size energy splitting. The splitting can be obtained directly from the unfolded Dirac equation, Eq.~\eqref{eq:unfolded_dirac}. Writing the two-component wavefunction as
\begin{equation}
\Phi(y)=
\begin{pmatrix}
\alpha(y)\\
\gamma(y)
\end{pmatrix},
\end{equation}
the Dirac equation in a region where the mass $m$ is constant can be rearranged into
\begin{equation}
\partial_y\Phi(y)=A(m)\Phi(y),
\qquad
A(m)=\frac{i}{v}
\begin{pmatrix}
E&-m\\
m&-E
\end{pmatrix}.
\end{equation}
Because $A(m)$ is constant within this region, the wavefunction at two points separated by a distance $L$ is related by
\begin{equation}
\Phi(y+L)=T(m)\Phi(y),
\qquad
T(m)=e^{A(m)L}.
\end{equation}
Thus, $T(m)$ is the transfer matrix that relates the wavefunction at the beginning of a constant-mass region to the wavefunction at its end. Since
\begin{equation}
A(m)^2=\kappa^2 I,
\qquad
\kappa=\frac{\sqrt{M_\Delta^2-E^2}}{v},
\end{equation}
for $m=\pm M_\Delta$, the exponential can be evaluated explicitly as
\begin{equation}
T(m)=\cosh(\kappa L)I+
\frac{\sinh(\kappa L)}{\kappa}A(m).
\end{equation}

In the unfolded problem, one half of the ring has mass $-M_\Delta$ and the other half has mass $+M_\Delta$. Therefore, after passing through both regions, the wavefunction is related to its initial value by the product $T(-M_\Delta)T(M_\Delta)$. At the same time, the unfolded field obeys the twisted boundary condition $\Phi(y+2L)=e^{-2ik_FL}\Phi(y)$. Combining these two statements requires $e^{-2ik_FL}$ to be an eigenvalue of the total transfer matrix. Since the transfer matrix has unit determinant, this condition can be written as
\begin{equation}
{\rm tr}\left[T(-M_\Delta)T(M_\Delta)\right]
=
2\cos(2k_FL).
\end{equation}
Evaluating the trace gives the exact quantization condition
\begin{equation}
E\,\sinh(\kappa L)=\pm v\kappa\,\sin(k_FL).
\label{eq:continuum_quantization}
\end{equation}

For a long chain, $L\gg\xi$, the lowest-energy solution satisfies $|E|\ll M_\Delta$. In this limit $\kappa\simeq M_\Delta/v=1/\xi$, where $\xi=v/M_\Delta$, and $\sinh(\kappa L)\simeq e^{L/\xi}/2$. Equation~\eqref{eq:continuum_quantization} then gives
\begin{equation}
E\simeq\pm2M_\Delta e^{-L/\xi}\sin(k_FL).
\end{equation}
The energy difference between the two lowest states is
\begin{equation}
E_{\rm split}(L,\mu)\simeq
2M_\Delta\,e^{-L/\xi}\,
\left|\sin(k_FL)\right|.
\label{eq:finite_mu_splitting}
\end{equation}
The factor $e^{-L/\xi}$ reflects the exponentially small overlap of the two edge modes, while the factor $\sin(k_FL)$ comes from the oscillatory part of their wavefunctions. Consequently, the splitting is exponentially suppressed with increasing system size but oscillates with $k_FL$. It vanishes whenever $k_FL$ is an integer multiple of $\pi$. For example, at $\mu=0$ one has $k_Fa_0=\pi/2$, and since $L=(N+1)a_0$, the splitting vanishes for every odd $N$.

In the lowest two-state subspace, this hybridization can be written in terms of the two Majorana operators as $H\simeq\frac{i}{2}E_{\rm split}\Gamma_L\Gamma_R$. Equivalently, the two edge Majoranas can be combined into the nonlocal fermion
\begin{equation}
d=\frac{\Gamma_L+i\Gamma_R}{2},
\qquad
H\simeq
E_{\rm split}
\left(d^\dagger d-\frac12\right).
\label{eq:nonlocal_fermion}
\end{equation}
The states with $d^\dagger d=0$ and $d^\dagger d=1$ have opposite fermion parity and differ in energy by $E_{\rm split}$. In the thermodynamic limit $L\rightarrow\infty$, the overlap of the two edge modes vanishes exponentially, so $E_{\rm split}\rightarrow0$ and the two states become degenerate.

As a check, we compare with the exact lattice Bogoliubov--de Gennes problem of Eq.~\eqref{eq:lattice_hamiltonian} at $U=0$. There, the zero-energy edge profile obeys $(t+\Delta)f_{j+1}+\mu f_j+(t-\Delta)f_{j-1}=0$, whose characteristic roots are
\begin{equation}
r_\pm=
\frac{-\mu\pm\sqrt{\mu^2-4(t^2-\Delta^2)}}{2(t+\Delta)}.
\label{eq:majorana_roots}
\end{equation}
Both roots lie inside the unit circle exactly when $|\mu|<2t$, which reproduces the topological interval. In the complex-root regime, $|r_\pm|=\sqrt{(t-\Delta)/(t+\Delta)}$, which gives the lattice decay length
\begin{equation}
\xi_{\rm lat}^{-1}=\frac{1}{2a_0}
\ln\left|\frac{t+\Delta}{t-\Delta}\right|,
\label{eq:exact_lattice_coherence_length}
\end{equation}
and reduces to the continuum value $\xi=ta_0/\Delta$ at weak pairing. Similarly, the phase of the roots approaches $k_F$. 

The finite-size bosonized construction of Ref.~\cite{chua2020majorana} corresponds to the $\mu=0$ case of the above analysis. At finite $\mu$, the twist $e^{-2ik_FL}$ in Eq.~\eqref{eq:twisted_bc} changes with $\mu$, and this is what produces the oscillating factor $\sin(k_FL)$ in the splitting.

\subsubsection{Interacting boundary sector for $K\neq1$}

We now return to arbitrary $K$. The boundary conditions, finite-size mode algebra, compact pair $(\theta_0,Q)$, and parity decomposition derived above remain valid. Within the topological superconducting regime, where the Umklapp coupling has flowed to weak coupling or is absent from the infrared theory, the relevant superconducting sector is
\begin{align}
H_{\rm SC}
&=\frac{v}{2\pi}\int_0^L\mathrm{d}x\left[K(\partial_x\theta)^2 +\frac{1}{K} (\partial_x\phi)^2\right]\nonumber\\
&\quad+\frac{\mu}{\pi}\int_0^L\mathrm{d}x\,\partial_x\phi + \frac{v y_\Delta}{2\pi a_0^2} \int_0^L\mathrm{d}x\,\cos[2\theta(x)].
\label{eq:finite_interacting_SG}
\end{align}
The chemical-potential term is the same zero-mode contribution derived above and shifts $\bar Q$ without changing the parity decomposition.

Writing
\begin{equation}
\theta(x)=\theta_0+\theta_{\rm osc}(x),
\end{equation}
the pairing vertex contains the zero-mode factors $e^{\pm2i\theta_0}$ and changes $Q$ by $\mp2$.  Hence
\begin{equation}
[H_{\rm SC},P]=0
\end{equation}
for arbitrary $K$.  The even and odd parity sectors are exact sectors of the interacting bosonized Hamiltonian.

The pairing operator has a scaling dimension
\begin{equation}
\Delta[\cos(2\theta)] = \frac{1}{K},
\end{equation}
and its weak-coupling flow is
\begin{equation}
\frac{dy_\Delta}{d\ell} = \left( 2-\frac{1}{K} \right)y_\Delta+\cdots.
\end{equation}
Together with the flow of $K$, this determines the superconducting
strong-coupling scale.  If $\ell_\ast$ is the scale at which the renormalized pairing becomes order unity, the interacting correlation length and mass are 
\begin{equation}
\xi(K)=a_0e^{\ell_\ast}, \quad M(K)\sim\frac{v}{\xi(K)}.
\label{eq:interacting_xi_mass}
\end{equation}
Away from the immediate BKT regime and when the running of $K$ is weak during the initial flow,
\begin{equation}
\xi(K) \sim a_0 |y_\Delta|^{-1/(2-1/K)}.
\label{eq:xi_K_scaling}
\end{equation}
Near the BKT separatrix, the coupled flow gives the essential divergence of $\xi$ discussed above.

At scales larger than $\xi(K)$, the pairing term flows to strong coupling and the bulk is gapped. Fermion parity remains exactly conserved, so the even and odd sectors do not mix. For a finite open chain, however, their lowest energies can differ by an exponentially small amount associated with the coupling of the two boundaries through the gapped bulk. We expect an exponentially small splitting of the form
\begin{equation}
\delta E_K(L,\mu) \sim A_K(L,\mu)e^{-L/\xi_{\rm eff}},
\label{eq:interacting_parity_splitting}
\end{equation}
where the overall amplitude and effective decay length are not fixed by the weak-coupling renormalization-group equations alone. The dependence of the prefactor on $L$ and $\mu$ can nevertheless be understood from the compact zero-mode sector. At strong coupling, the pairing term pins $\theta_0$ at the two parity-related minima, while the offset charge $\bar Q$ controls the relative phase between tunneling paths connecting them. The interference between these paths gives
\begin{equation}
\delta E_K(L,\mu) \sim \Lambda_K e^{-L/\xi_{\rm eff}}\left|\cos(\pi\bar Q)\right|.
\label{eq:interacting_zero_mode_splitting}
\end{equation}
The splitting vanishes when $\bar Q$ is half-integer. At $K=1$, one has $\pi\bar Q=k_FL-\pi/2$ to linear order in $\mu$, so that $|\cos(\pi\bar Q)|=|\sin(k_FL)|$, reproducing Eq.~\eqref{eq:finite_mu_splitting}. For $K\neq1$, the interference factor follows from the compact zero-mode structure, while $\Lambda_K$ and $\xi_{\rm eff}$ remain interaction dependent.

The corresponding operator within the lowest two-state subspace can also
be written directly in bosonic language. Let $\mathcal P_{\rm low}$ project onto the lowest state in each parity sector.  Since $e^{\pm i\theta_0}$ changes $Q$ by one,
\begin{align}
X_{\rm low} &= \mathcal N_X \mathcal P_{\rm low} \left(
e^{i\theta_0} + e^{-i\theta_0} \right) \mathcal P_{\rm low}
\label{eq:projected_parity_changer}
\end{align}
obeys
\begin{equation}
\{P,X_{\rm low}\}=0.
\end{equation}
When its off-diagonal matrix element is nonzero, the choice $\mathcal N_X=|\langle +|\,2\cos\theta_0\,|-\rangle|^{-1}$, where $|\pm\rangle$ are the lowest states of the two parity sectors, gives
\begin{equation}
X_{\rm low}^2=\mathcal P_{\rm low}.
\end{equation}
This operator specifies the parity-changing action after the massive bulk states have been projected out.  It should be distinguished from the full spatial edge operator.  At $K=1$ the latter is given by Eq.~\eqref{eq:continuum_majorana_L}; for generic $K$ it acquires
interaction-dependent many-body dressing.

The interacting bosonic theory determines the parity-resolved finite-size Hilbert space and the superconducting correlation length. It also motivates the exponentially small finite-size splitting in the topological phase, although it does not by itself provide an explicit spatially resolved interacting Majorana operator. At $K=1$, the exact continuum construction is given above. For generic $K$, the boundary operators are expected to become quasi-local, interaction-dependent dressings of the free-fermion edge operators as long as fermion parity is preserved and the many-body bulk gap remains open.

\subsection{Dual symmetry-breaking representation of the topological phase}\label{dual-ssb}

The topological superconducting phase is naturally characterized in the original fermionic representation by its fermion-parity structure and Majorana boundary excitations. The same continuum bulk Hamiltonian admits a complementary formal representation in which the gapped phase
is characterized by conventional discrete symmetry breaking. To make this correspondence explicit, we first perform a canonical transformation of the bosonized fields and subsequently refermionize the transformed theory in terms of a new set of dual fermions. The
resulting fermions define a different local operator representation from the microscopic Kitaev fermions.

In the superconducting regime the relevant bulk sector is Eq.~\eqref{eq:effective_hamiltonian} with $y_U$ irrelevant, i.e. the
sine-Gordon theory in $\theta$ whose pairing perturbation flows to strong
coupling for $K>1/2$.

We introduce a new canonical pair $(\varphi,\vartheta)$ through
\begin{equation}
\varphi=\frac{\theta}{2}, \quad \vartheta=2\phi.
\label{eq:SSB_duality}
\end{equation}
The continuous transformation maps the two generators of the original
identification lattice according to
\begin{align}
(\phi+\pi,\theta+\pi) &\mapsto \left(\varphi+\frac{\pi}{2}, \vartheta+2\pi\right),\nonumber\\
(\phi,\theta+2\pi)&\mapsto(\varphi+\pi,\vartheta).
\label{eq:image_lattice}
\end{align}
The local dual fermions are defined with the compactification
\begin{equation}
(\varphi,\vartheta)\sim(\varphi+\pi,\vartheta+\pi), \quad (\varphi,\vartheta)\sim(\varphi,\vartheta+2\pi).
\label{eq:dual_compactification}
\end{equation}
The image of the original lattice is distinct from the compactification lattice used to define the dual local operators.  After this change of local operator algebra, the image shifts are represented as global transformations of the dual Hilbert space rather than as dual field identifications. Since $\vartheta\rightarrow\vartheta+2\pi$ is a dual identification, the first image shift acts as $\mathcal S:\varphi\rightarrow\varphi+\pi/2$, while the second acts as $\mathcal S^2:\varphi\rightarrow\varphi+\pi$.

Using
\begin{equation}
[\phi(x),\partial_y\theta(y)] = i\pi\delta(x-y),
\end{equation}
and equivalently
\begin{equation}
[\theta(x),\partial_y\phi(y)] = i\pi\delta(x-y),
\end{equation}
one finds
\begin{align}
[\varphi(x),\partial_y\vartheta(y)] &= \left[ \frac{\theta(x)}{2}, \partial_y(2\phi(y))
\right] \nonumber\\
&= i\pi\delta(x-y).
\end{align}
Thus $(\varphi,\vartheta)$ form a canonically conjugate pair in the same normalization as $(\phi,\theta)$.

The inverse transformation is
\begin{equation}
\theta=2\varphi, \quad \phi=\frac{\vartheta}{2}.
\end{equation}
Substitution into the Gaussian Hamiltonian gives
\begin{align}
H_{LL} &= \frac{v}{2\pi} \int \mathrm{d}x \left[ 4K(\partial_x\varphi)^2 +
\frac{1}{4K}(\partial_x\vartheta)^2 \right] \nonumber\\
&= \frac{v}{2\pi} \int \mathrm{d}x \left[ K_{\rm d}(\partial_x\vartheta)^2 +
\frac{1}{K_{\rm d}}(\partial_x\varphi)^2 \right],
\end{align}
where
\begin{equation}
K_{\rm d} = \frac{1}{4K}.
\label{eq:dual_K}
\end{equation}
At the same time,
\begin{equation}
\cos(2\theta) = \cos(4\varphi),
\label{eq:pairing_to_dual_umklapp}
\end{equation}
and the superconducting bulk theory becomes
\begin{align}
H_{\rm d} &= \frac{v}{2\pi}\int\mathrm{d}x \left[ K_{\rm d}(\partial_x\vartheta)^2 +\frac{1}{K_{\rm d}}(\partial_x\varphi)^2 \right]
\nonumber\\
&\quad+\frac{\mu}{2\pi} \int\mathrm{d}x\,\partial_x\vartheta + \frac{v y_\Delta}{2\pi a_0^2} \int\mathrm{d}x\,\cos(4\varphi).
\label{eq:dual_SG}
\end{align}
At $\mu=0$, Eq.~\eqref{eq:dual_SG} has the same sine-Gordon form as the commensurate Umklapp theory of Sec.~\ref{sec:mott}, with $K\rightarrow K_{\rm d}=1/(4K)$ and $y_U\rightarrow y_\Delta$. For finite $\mu$, the additional total-derivative term shifts the dual winding sectors but is invariant under the dual symmetry introduced below. The transformation is self-dual at $K=K_{\rm d}=1/2$.

The transformation preserves the scaling dimension of the gapping operator. In the original variables,
\begin{equation}
\Delta[\cos(2\theta)] = \frac{1}{K},
\end{equation}
whereas in the dual variables
\begin{equation}
\Delta[\cos(4\varphi)] = 4K_{\rm d} = \frac{1}{K}.
\end{equation}
Thus the superconducting instability and the dual locking instability become relevant in precisely the same regime.

The canonical transformation specifies the map between the bosonic phase spaces, but the duality also changes the notion of locality. To make this explicit, we now refermionize the transformed theory by introducing a new set of fermionic fields $\chi_{R/L}$ according to
the standard bosonization dictionary
\begin{equation}
\chi_r(x) = \frac{1}{\sqrt{2\pi a_0}} \exp\left[ -i\left( r\varphi(x)-\vartheta(x)
\right) \right], ~~ r=\pm1.
\label{eq:dual_fermions}
\end{equation}
The dual fermions $\chi_{R/L}$ are not the microscopic Kitaev fermions $\psi_{R/L}$; rather, they define the elementary local fermionic operators of the dual representation. For $K_{\rm d}\neq1$, the $\chi$ fermions provide an interacting fermionic representation of the Luttinger liquid, and they are free fermions only at $K_{\rm d}=1$.

At the level of the continuous fields, Eq.~\eqref{eq:SSB_duality} is an
involution so  applying it twice returns $(\vartheta/2,2\varphi)=(\phi,\theta)$. Correspondingly, the parameter map is $K\mapsto1/(4K)\mapsto K$, with the self-dual value $K=1/2$ as its fixed point. The continuous field transformations have this symmetry, while the two compactified local operator algebras assign different notions of locality. Using Eq.~\eqref{eq:SSB_duality}, the dual fermions are represented in the original bosonic variables as
\begin{equation}
\chi_r(x) = \frac{1}{\sqrt{2\pi a_0}} \exp\left[ -i\left(
\frac{r}{2}\theta(x)-2\phi(x) \right) \right],
\label{eq:dual_fermions_original}
\end{equation}
whereas the original fermions become, in terms of the dual fields,
\begin{equation}
\psi_r(x) = \frac{1}{\sqrt{2\pi a_0}} \exp\left[ -i\left(\frac{r} {2}\vartheta(x)-2\varphi(x)\right)\right].
\label{eq:original_fermions_dual}
\end{equation}
The difference between the representations lies in the charge labels carried relative to the identification lattice that defines each theory. Suppressing normalization constants, in terms of the vertex operators $V_{n,m}=e^{i(n\phi+m\theta)}$ of Sec.~\ref{sec:RG},
\begin{equation}
\psi_r=V_{-r,\,+1},\quad \chi_r=V_{+2,\,-r/2},
\label{eq:vertex_labels}
\end{equation}
so that $\chi_r$ carries half-integer $\theta$  charge. The half vertex $e^{\mp i\theta/2}$ is not single-valued under Eq.~\eqref{eq:identification_lattice} and is not an operator of the original algebra. Under the corresponding exchange of the two descriptions, $\psi_r$ stands in exactly the same relation to Eq.~\eqref{eq:dual_compactification}. The two fermion families are mutually nonlocal, and the continuum theory alone does not privilege either. It
is the microscopic lattice Hamiltonian Eq.~\eqref{eq:lattice_hamiltonian} that selects $\psi_{R/L}$ as the elementary local fermions.

The same statement can be made quantitatively through the scaling dimensions. Evaluated in the original variables,
\begin{equation}
\Delta[\chi_r]=\frac14\left(4K+\frac{1}{4K}\right)=K+\frac{1}{16K},
\end{equation}
which coincides with $\tfrac14(K_{\rm d}+1/K_{\rm d})$ evaluated in the dual theory, as it must, and reaches the free-fermion value $1/2$ only at $K=1/4$,
i.e. $K_{\rm d}=1$. Conversely $\Delta[\psi_r]=\tfrac14(K+1/K)$ equals $1/2$
only at $K=1$, where $\Delta[\chi_r]=17/16$. Since $K_{\rm d}=1/(4K)$, the two fermion families can never be simultaneously free. Moreover, throughout the superconducting relevance regime $K>1/2$, one has $K_{\rm d}<1/2$, so the dual $\chi$ fermions are never free within the topological superconducting regime considered here. Their free-fermion point, $K_{\rm d}=1$ or equivalently $K=1/4$, lies outside this regime and, at commensurability, on the charge-ordered side of the weak-coupling phase diagram. This exchange of locality is what allows the same continuum bulk Hamiltonian to admit both a topological and a symmetry-breaking representation. It does not by itself establish an exact microscopic lattice duality.

The symmetry-breaking character of the dual representation can now be stated directly in terms of local operators of the $\chi$ fermions. The locking perturbation $\cos(4\varphi)$ is invariant under
\begin{equation}
\mathcal{S}:\quad \varphi\longrightarrow\varphi+\frac{\pi}{2},
\label{eq:dual_symmetry}
\end{equation}
since
\begin{equation}
\cos\left[4\left(\varphi+\frac{\pi}{2}\right)\right] = \cos(4\varphi).
\end{equation}

Using the dual fermion dictionary, this transformation acts as
\begin{equation}
\mathcal{S}:\quad \chi_R\longrightarrow-i\chi_R, \quad \chi_L\longrightarrow+i\chi_L.
\label{eq:dual_symmetry_fermions}
\end{equation}
Hence,
\begin{equation}
\mathcal S^2=P_\chi,\quad P_\chi=(-1)^{F_\chi},
\end{equation}
where $P_\chi$ denotes the parity of the dual fermions. Moreover, $\mathcal S^4=1$, since four applications of $\mathcal S$ shift $\varphi\rightarrow\varphi+2\pi$, and
$(2\pi,0)=2(\pi,\pi)-(0,2\pi)$ belongs to the compactification lattice in  Eq.~\eqref{eq:dual_compactification}. Thus $\mathcal S$ is a fermionic $\mathbb Z_4^f$ symmetry, with $\mathcal S^2=P_\chi$, whose action on parity-even local bilinears reduces to an ordinary $\mathbb Z_2$ symmetry.

In the original variables, the same transformation is
$\theta\rightarrow\theta+\pi$, which is the action of the microscopic fermion parity $P_\psi=(-1)^F$. Under the duality, this microscopic parity is represented by $\mathcal S$, whereas $P_\chi=\mathcal S^2$ is the fermion parity of the dual fermions. The order parameter introduced below is odd under $\mathcal S$ but even under $P_\chi$. In the dual representation, it is an ordinary fermion-parity-even density-wave bilinear, and its expectation value is an admissible order parameter. In the microscopic representation the same operator, Eq.~\eqref{eq:dual_order_original}, is odd under $P_\psi=\mathcal S$, hence forbidden from acquiring an expectation value by the fermion-parity superselection rule, quite apart from being nonlocal. In the original variables $\mathcal S^2$ acts as $\theta\to\theta+2\pi$, which is an identification of Eq.~\eqref{eq:identification_lattice}, so $P_\chi$ acts trivially on all operators built from $\psi_{R/L}$. The duality thus maps the microscopic fermion parity $P_\psi$ onto $\mathcal S$, while the dual fermion parity $P_\chi=\mathcal S^2$ is invisible to the microscopic local operators, and this exchange is what converts a superselected, unbreakable symmetry into an ordinary, breakable one. The two descriptions use different notions of locality and  different local operator algebras.

For definiteness, we choose the sign convention $y_\Delta<0$, so that the strong-coupling minima occur at
\begin{equation}
\varphi=\frac{n\pi}{2},\quad n\in\mathbb Z.
\label{eq:dual_minima}
\end{equation}
Within the interval $0\leq\varphi<2\pi$, these give four coordinate representatives. However, the identification lattice relates
\begin{equation}
(0,\vartheta_0)\sim(\pi,\vartheta_0+\pi),
\quad \left(\frac{\pi}{2},\vartheta_0\right)
\sim
\left(\frac{3\pi}{2},\vartheta_0+\pi\right).
\label{eq:dual_minima_identifications}
\end{equation}
Thus there are only two inequivalent vacua, which may be represented by
\begin{equation}
\varphi=0 \quad\text{and}\quad \varphi=\frac{\pi}{2}.
\label{eq:dual_vacua}
\end{equation}
The symmetry $\mathcal S:\varphi\rightarrow\varphi+\pi/2$ exchanges these two vacua, while $\mathcal S^2=P_\chi$ acts within each vacuum sector.

In the dual fermionic representation, the local $2k_F$ density-wave bilinear is
\begin{equation}
\chi_R^\dagger\chi_L+\mathrm{H.c.} \propto \cos(2\varphi).
\end{equation}
We define the dual order parameter as
\begin{equation}
O_{\rm SSB}(x)\sim\cos[2\varphi(x)].
\label{eq:dual_order_parameter}
\end{equation}
In a finite system with an exact $\mathcal S$ symmetry, the exact symmetry  eigenstates satisfy
\begin{equation}
\langle O_{\rm SSB}\rangle=0.
\end{equation}
A nonzero expectation value is understood in the thermodynamic limit, in a symmetry-broken linear combination of ground states, or in the presence of an infinitesimal symmetry-breaking field.

Under Eq.~\eqref{eq:dual_symmetry},
\begin{equation}
O_{\rm SSB}\longrightarrow-O_{\rm SSB},
\end{equation}
whereas the locking perturbation $\cos(4\varphi)$ remains invariant. When the locking perturbation flows to strong coupling, $\mathcal S$ is spontaneously broken down to $P_\chi$. At the semiclassical minima, the order parameter has opposite signs,
\begin{align}
\cos(2\varphi)&=+1 \quad (\varphi=0),\\
\cos(2\varphi)&=-1 \quad \left(\varphi=\tfrac{\pi}{2}\right).
\end{align}
In the interacting quantum theory, the magnitude of the expectation value is generally renormalized, but the two vacua are distinguished by the sign of $O_{\rm SSB}$, exactly as the two period-two charge-ordered states of Sec.~\ref{sec:mott} are distinguished by the sign of $m_{\rm CDW}$. If instead $y_\Delta>0$ the minima are shifted to $\varphi=\pi/4$ and $3\pi/4$, and the order parameter is $\sin(2\varphi)$ rather than $\cos(2\varphi)$.

The counting can also be carried out in the original variables.  The shift $\theta\rightarrow\theta+\pi$ is an identification only when accompanied by $\phi\rightarrow\phi+\pi$.  Since $\phi$ is not pinned in the superconducting phase, the minima of $\cos(2\theta)$ are all equivalent under the original compactification and represent a single inequivalent bulk vacuum.  The microscopic parity $P_\psi$ acts as $(\phi,\theta)\rightarrow(\phi,\theta+\pi)$, which is equivalent under Eq.~\eqref{eq:identification_lattice} to $(\phi,\theta)\rightarrow(\phi-\pi,\theta)$.  It shifts the disordered field rather than the pinned field and is represented on the finite-size zero mode by
\begin{equation}
P_\psi=e^{i\pi Q}.
\end{equation}
The topological superconducting phase has no bulk symmetry breaking or local order parameter.  In an open chain, its asymptotic two-state low-energy structure is carried by the boundary parity sector, whereas the dual description has two minima exchanged by $\mathcal S$ and
a local symmetry-breaking order parameter.

The same relation illustrates explicitly how locality is exchanged between the two descriptions. From Eq.~\eqref{eq:SSB_duality},
\begin{equation}
O_{\rm SSB}(x)\sim \cos\left[\theta(x)-\alpha_\Delta\right],
\label{eq:dual_order_original}
\end{equation}
where $\alpha_\Delta=0$ for $y_\Delta<0$ and $\alpha_\Delta=\pi/2$ for $y_\Delta>0$. The order parameter is local in the dual fermionic representation but generally nonlocal when expressed in terms of the microscopic Kitaev fermions; i.e., the left-hand side is a local bilinear of the dual fermions $\chi_{R/L}$, while the right-hand side is not a local fermion bilinear of the microscopic Kitaev variables. Conversely,
Eq.~\eqref{eq:original_fermions_dual} shows that the microscopic Kitaev fermions themselves become nontrivial vertex operators in the dual representation. The topological and symmetry-breaking descriptions  employ different local operator algebras even though their continuum Hamiltonians are related by the canonical field transformation.

This change of compactification is analogous to the orbifold or gauging
description of fermion-parity dualities. The local operator algebra and its
allowed winding sectors are changed together.  Here we use this language as
a characterization of the continuum construction. Since the image of the original identification lattice and the dual compactification lattice do not contain one another, the correspondence is not a simple orbifold of one theory by a subgroup, and a complete gauging
construction would require an explicit treatment of all twisted sectors.

This exchange of locality is closely related to the SPT--SSB correspondence discussed in Ref.~\cite{tang2026topological}. In that work, a nonlocal transformation relates fermionic representations with conventional symmetry breaking and symmetry-protected topological boundary structure. The present continuum construction follows the same exchange of locality, but it should be viewed as an analogy and continuum realization rather than as a proof of an exact microscopic mapping.

The lattice Hamiltonian of Ref.~\cite{tang2026topological} belongs to a special parameter family whose relations between hopping, pairing, and interaction permit an exact solution. The resulting topological phases, however, are not restricted to a fine-tuned point and remain stable under symmetry-preserving perturbations that do not close the bulk gap. The continuum construction above follows the same exchange of locality where the dual fermions $\chi_{R/L}$ provide a local symmetry-breaking description of the
superconducting bulk sector, whereas the microscopic Kitaev fermions $\psi_{R/L}$ provide the topological description with Majorana boundary physics. In this sense, the canonical transformation in Eq.~\eqref{eq:SSB_duality} provides a continuum realization of the SPT--SSB correspondence.

\section{Conclusion}\label{conclusion}

We have developed a field-theoretic description of the interacting Kitaev chain at finite chemical potential and compared its predictions with DMRG. The bulk physics is controlled by two competing locking terms. Superconducting pairing pins $\theta$ through $\cos(2\theta)$, while the Umklapp term $\cos(4\phi)$ favors commensurate charge order by pinning $\phi$. Since the two fields are canonically conjugate, both locking mechanisms cannot operate simultaneously. A finite chemical potential couples to the topological charge $Q_\phi$ and weakens the commensurate locking by introducing solitons. Away from commensurability, an incommensurate Luttinger liquid then appears between the charge-density-wave Mott phase and the topological superconductor. At sufficiently large chemical potential the system instead enters a trivial band insulator.

The phase boundaries are governed by different critical theories. At $\mu=0$, the topological and charge-ordered phases meet at the exactly known XYZ-chain point $U_c=(t+\Delta)/2$. This transition lies on the eight-vertex critical line, where $\nu=\pi/(2\rho)$ and $\beta=(\pi-\rho)/(4\rho)$ obey $4\beta=2\nu-1$. At finite doping the direct transition is replaced by an incommensurate region. The transition out of the commensurate Mott phase is of Pokrovsky--Talapov type where the soliton density has a square-root onset, $z=2$, and $K$ approaches $1/4$. On entering the topological superconducting phase, the transition is instead of Berezinskii--Kosterlitz--Thouless type and occurs at $K=1/2$. The topological phase also has a boundary with the trivial insulator at larger chemical potential. This is an Ising transition with a single massless Majorana mode, $z=1$, and $c=1/2$. Our DMRG calculations reproduce the corresponding phase structure.

We have also studied the Majorana edge modes directly in the continuum theory. Their description differs from the familiar boundary soliton of the SSH chain. There, the bulk pins $\phi$, and the Dirichlet boundary condition can force a charged half-kink near the edge~\cite{jin2023bosonization}. In the Kitaev chain the pairing term pins $\theta$, which obeys a Neumann boundary condition at a hard wall. The Majorana mode does not appear as a classical density kink. Close to the topological--trivial transition, it is the zero mode bound to a sign change of the band-edge Ising mass. In the absence of interactions, the open-chain continuum theory can be solved explicitly and gives one exponentially localized Majorana mode at each end. At $K=1$ their wavefunctions oscillate with $k_F$, and their overlap produces the finite-size splitting $2M_\Delta e^{-L/\xi}|\sin(k_FL)|$. The lattice calculation approaches the same result in the continuum limit.

For finite interactions, the compact zero modes of the bosonized theory become important. Keeping the conjugate pair $(\theta_0,Q)$ separates the open-chain Hilbert space into sectors of opposite fermion parity. These sectors remain distinct in the interacting theory, while the difference between their lowest energies vanishes exponentially with system size. The dependence of this splitting on the charge offset $\bar Q$ follows from the compact zero-mode sector. Thus, even though the interacting bosonized theory does not give an explicit spatial form for the dressed Majorana operator, it retains the parity structure associated with the topological phase.

We finally considered a different fermionic representation of the same low-energy bulk theory. Under the canonical transformation $(\varphi,\vartheta)=(\theta/2,2\phi)$, the superconducting term becomes $\cos(4\varphi)$ and $K$ is mapped to $K_{\rm d}=1/(4K)$. Refermionizing these fields defines dual fermions with a different notion of locality from the microscopic Kitaev fermions. Microscopic fermion parity becomes the generator of a fermionic $\mathbb Z_4^f$ symmetry in the dual representation, with the dual fermion parity given by its square. When $\varphi$ is pinned, this symmetry is broken down to the dual fermion parity, and $\cos(2\varphi)$ acts as a local order parameter. The phase that is topological in terms of the microscopic Kitaev fermions is symmetry-breaking in the dual fermionic variables. This is the continuum form of the SPT--SSB correspondence discussed in Ref.~\cite{tang2026topological}.

\textit{Acknowledgments}: This work is supported by the Swiss National Science Foundation under grant number 200020-219400.

\bibliography{ref}

@article{kitaev2001unpaired,
  author  = {Kitaev, A. Yu.},
  title   = {Unpaired Majorana fermions in quantum wires},
  journal = {Physics-Uspekhi},
  volume  = {44},
  number  = {10S},
  pages   = {131--136},
  year    = {2001},
  doi     = {10.1070/1063-7869/44/10S/S29},
  url     = {https://doi.org/10.1070/1063-7869/44/10S/S29}
}

@article{katsura2015exact,
  author  = {Katsura, Hosho and Schuricht, Dirk and Takahashi, Masahiro},
  title   = {Exact ground states and topological order in interacting {Kitaev/Majorana} chains},
  journal = {Physical Review B},
  volume  = {92},
  number  = {11},
  pages   = {115137},
  year    = {2015},
  doi     = {10.1103/PhysRevB.92.115137},
  url     = {https://doi.org/10.1103/PhysRevB.92.115137}
}

@article{chepiga2023eight,
  author  = {Chepiga, Natalia and Mila, Fr{\'e}d{\'e}ric},
  title   = {Eight-vertex criticality in the interacting {Kitaev} chain},
  journal = {Physical Review B},
  volume  = {107},
  number  = {8},
  pages   = {L081106},
  year    = {2023},
  doi     = {10.1103/PhysRevB.107.L081106},
  url     = {https://doi.org/10.1103/PhysRevB.107.L081106}
}

@article{lieb1961two,
  author  = {Lieb, Elliott and Schultz, Theodore and Mattis, Daniel},
  title   = {Two soluble models of an antiferromagnetic chain},
  journal = {Annals of Physics},
  volume  = {16},
  number  = {3},
  pages   = {407--466},
  year    = {1961},
  doi     = {10.1016/0003-4916(61)90115-4},
  url     = {https://doi.org/10.1016/0003-4916(61)90115-4}
}

@book{fradkin2013field,
  author    = {Fradkin, Eduardo},
  title     = {Field Theories of Condensed Matter Physics},
  edition   = {2},
  publisher = {Cambridge University Press},
  year      = {2013},
  doi       = {10.1017/CBO9781139015509},
  url       = {https://doi.org/10.1017/CBO9781139015509},
  isbn      = {978-0-521-76444-5}
}

@book{giamarchi2003quantum,
  author    = {Giamarchi, Thierry},
  title     = {Quantum Physics in One Dimension},
  series    = {International Series of Monographs on Physics},
  volume    = {121},
  publisher = {Clarendon Press},
  address   = {Oxford},
  year      = {2003},
  isbn      = {9780198525004},
  doi       = {10.1093/acprof:oso/9780198525004.001.0001},
  url       = {https://doi.org/10.1093/acprof:oso/9780198525004.001.0001}
}

@book{gogolin2004bosonization,
  author    = {Gogolin, Alexander O. and Nersesyan, Alexander A. and Tsvelik, Alexei M.},
  title     = {Bosonization and Strongly Correlated Systems},
  publisher = {Cambridge University Press},
  year      = {2004},
  isbn      = {978-0-521-61719-2},
  url       = {https://www.cambridge.org/9780521617192}
}

@article{berezinskii1971destruction,
  author  = {Berezinskii, V. L.},
  title   = {Destruction of Long-Range Order in One-Dimensional and Two-Dimensional Systems Having a Continuous Symmetry Group. I. Classical Systems},
  journal = {Soviet Physics JETP},
  volume  = {32},
  number  = {3},
  pages   = {493--500},
  year    = {1971},
  url     = {https://www.jetp.ras.ru/cgi-bin/dn/e_032_03_0493.pdf}
}

@article{kosterlitz1973ordering,
  author  = {Kosterlitz, J. M. and Thouless, D. J.},
  title   = {Ordering, Metastability and Phase Transitions in Two-Dimensional Systems},
  journal = {Journal of Physics C: Solid State Physics},
  volume  = {6},
  number  = {7},
  pages   = {1181--1203},
  year    = {1973},
  doi     = {10.1088/0022-3719/6/7/010},
  url     = {https://doi.org/10.1088/0022-3719/6/7/010}
}

@article{pokrovsky1979ground,
  title   = {Ground State, Spectrum, and Phase Diagram of Two-Dimensional Incommensurate Crystals},
  author  = {Pokrovsky, V. L. and Talapov, A. L.},
  journal = {Physical Review Letters},
  volume  = {42},
  number  = {1},
  pages   = {65--67},
  year    = {1979},
  doi     = {10.1103/PhysRevLett.42.65},
  url     = {https://doi.org/10.1103/PhysRevLett.42.65},
  publisher = {American Physical Society}
}

@article{zamolodchikov1979factorized,
  title     = {Factorized S-Matrices in Two Dimensions as the Exact Solutions of Certain Relativistic Quantum Field Theory Models},
  author    = {Zamolodchikov, Alexander B. and Zamolodchikov, Alexey B.},
  journal   = {Annals of Physics},
  volume    = {120},
  number    = {2},
  pages     = {253--291},
  year      = {1979},
  doi       = {10.1016/0003-4916(79)90391-9},
  url       = {https://doi.org/10.1016/0003-4916(79)90391-9},
  publisher = {Academic Press}
}

@article{jackiw1976solitons,
  title     = {Solitons with Fermion Number $1/2$},
  author    = {Jackiw, Roman and Rebbi, Claudio},
  journal   = {Physical Review D},
  volume    = {13},
  number    = {12},
  pages     = {3398--3409},
  year      = {1976},
  doi       = {10.1103/PhysRevD.13.3398},
  url       = {https://doi.org/10.1103/PhysRevD.13.3398},
  publisher = {American Physical Society}
}

@article{calabrese2004entanglement,
  title     = {Entanglement Entropy and Quantum Field Theory},
  author    = {Calabrese, Pasquale and Cardy, John},
  journal   = {Journal of Statistical Mechanics: Theory and Experiment},
  volume    = {2004},
  number    = {06},
  pages     = {P06002},
  year      = {2004},
  doi       = {10.1088/1742-5468/2004/06/P06002},
  url       = {https://doi.org/10.1088/1742-5468/2004/06/P06002},
  publisher = {IOP Publishing}
}

@article{coleman1975quantum,
  title     = {Quantum Sine-Gordon Equation as the Massive Thirring Model},
  author    = {Coleman, Sidney},
  journal   = {Physical Review D},
  volume    = {11},
  number    = {8},
  pages     = {2088--2097},
  year      = {1975},
  doi       = {10.1103/PhysRevD.11.2088},
  url       = {https://doi.org/10.1103/PhysRevD.11.2088},
  publisher = {American Physical Society}
}

@article{bergknoff1979structure,
  title     = {Structure and Solution of the Massive Thirring Model},
  author    = {Bergknoff, H. and Thacker, H. B.},
  journal   = {Physical Review D},
  volume    = {19},
  number    = {12},
  pages     = {3666--3681},
  year      = {1979},
  doi       = {10.1103/PhysRevD.19.3666},
  url       = {https://doi.org/10.1103/PhysRevD.19.3666},
  publisher = {American Physical Society}
}

@article{iwabuchi1995commensurate,
  title     = {Commensurate-Incommensurate Phase Transition in One-Dimensional Quantum Sine-Gordon Model Based on a Lattice Massive Thirring Model},
  author    = {Iwabuchi, Shuichi},
  journal   = {Zeitschrift f{\"u}r Physik B: Condensed Matter},
  volume    = {97},
  number    = {3},
  pages     = {453--458},
  year      = {1995},
  doi       = {10.1007/BF01317228},
  url       = {https://doi.org/10.1007/BF01317228},
  publisher = {Springer}
}

@article{jin2023bosonization,
  title     = {Bosonization of the Interacting {Su-Schrieffer-Heeger} Model},
  author    = {Jin, Tony and Ruggiero, Paola and Giamarchi, Thierry},
  journal   = {Physical Review B},
  volume    = {107},
  number    = {20},
  pages     = {L201111},
  year      = {2023},
  doi       = {10.1103/PhysRevB.107.L201111},
  url       = {https://doi.org/10.1103/PhysRevB.107.L201111},
  publisher = {American Physical Society}
}

@article{zamolodchikov1995mass,
  title     = {Mass Scale in the Sine--Gordon Model and Its Reductions},
  author    = {Zamolodchikov, Al. B.},
  journal   = {International Journal of Modern Physics A},
  volume    = {10},
  number    = {08},
  pages     = {1125--1150},
  year      = {1995},
  doi       = {10.1142/S0217751X9500053X},
  url       = {https://doi.org/10.1142/S0217751X9500053X},
  publisher = {World Scientific}
}

@article{tang2026topological,
  title     = {On the Topological Dual of the {XXZ} Spin Chain},
  author    = {Tang, Yicheng and Kattel, Pradip and Andrei, Natan},
  journal   = {Physical Review B},
  volume    = {113},
  number    = {4},
  pages     = {L041113},
  year      = {2026},
  doi       = {10.1103/blkg-3t1v},
  url       = {https://doi.org/10.1103/blkg-3t1v},
  publisher = {American Physical Society}
}

@article{stoudenmire2011interaction,
  title     = {Interaction Effects in Topological Superconducting Wires Supporting Majorana Fermions},
  author    = {Stoudenmire, E. M. and Alicea, Jason and Starykh, Oleg A. and Fisher, Matthew P. A.},
  journal   = {Physical Review B},
  volume    = {84},
  number    = {1},
  pages     = {014503},
  year      = {2011},
  doi       = {10.1103/PhysRevB.84.014503},
  url       = {https://doi.org/10.1103/PhysRevB.84.014503},
  publisher = {American Physical Society}
}

@article{miao2018majorana,
  title     = {Majorana Zero Modes and Long Range Edge Correlation in Interacting {Kitaev} Chains: Analytic Solutions and Density-Matrix-Renormalization-Group Study},
  author    = {Miao, Jian-Jian and Jin, Hui-Ke and Zhang, Fu-Chun and Zhou, Yi},
  journal   = {Scientific Reports},
  volume    = {8},
  number    = {1},
  pages     = {488},
  year      = {2018},
  doi       = {10.1038/s41598-017-17699-y},
  url       = {https://doi.org/10.1038/s41598-017-17699-y},
  publisher = {Springer Nature}
}

@article{herviou2016phase,
  title     = {Phase Diagram and Entanglement of Two Interacting Topological {Kitaev} Chains},
  author    = {Herviou, Lo{\"i}c and Mora, Christophe and Le Hur, Karyn},
  journal   = {Physical Review B},
  volume    = {93},
  number    = {16},
  pages     = {165142},
  year      = {2016},
  doi       = {10.1103/PhysRevB.93.165142},
  url       = {https://doi.org/10.1103/PhysRevB.93.165142},
  publisher = {American Physical Society}
}

@article{alicea2012new,
  title     = {New Directions in the Pursuit of Majorana Fermions in Solid State Systems},
  author    = {Alicea, Jason},
  journal   = {Reports on Progress in Physics},
  volume    = {75},
  number    = {7},
  pages     = {076501},
  year      = {2012},
  doi       = {10.1088/0034-4885/75/7/076501},
  url       = {https://doi.org/10.1088/0034-4885/75/7/076501},
  publisher = {IOP Publishing}
}

@article{mourik2012signatures,
  title     = {Signatures of Majorana Fermions in Hybrid Superconductor-Semiconductor Nanowire Devices},
  author    = {Mourik, Vincent and Zuo, Kun and Frolov, Sergey M. and Plissard, S. R. and Bakkers, Erik P. A. M. and Kouwenhoven, Leo P.},
  journal   = {Science},
  volume    = {336},
  number    = {6084},
  pages     = {1003--1007},
  year      = {2012},
  doi       = {10.1126/science.1222360},
  url       = {https://doi.org/10.1126/science.1222360},
  publisher = {American Association for the Advancement of Science}
}

@article{das2012zero,
  title     = {Zero-Bias Peaks and Splitting in an {Al--InAs} Nanowire Topological Superconductor as a Signature of Majorana Fermions},
  author    = {Das, Anindya and Ronen, Yuval and Most, Yonatan and Oreg, Yuval and Heiblum, Moty and Shtrikman, Hadas},
  journal   = {Nature Physics},
  volume    = {8},
  number    = {12},
  pages     = {887--895},
  year      = {2012},
  doi       = {10.1038/nphys2479},
  url       = {https://doi.org/10.1038/nphys2479},
  publisher = {Nature Publishing Group}
}

@article{nadj2014observation,
  title     = {Observation of Majorana Fermions in Ferromagnetic Atomic Chains on a Superconductor},
  author    = {Nadj-Perge, Stevan and Drozdov, Ilya K. and Li, Jian and Chen, Hua and Jeon, Sangjun and Seo, Jungpil and MacDonald, Allan H. and Bernevig, B. Andrei and Yazdani, Ali},
  journal   = {Science},
  volume    = {346},
  number    = {6209},
  pages     = {602--607},
  year      = {2014},
  doi       = {10.1126/science.1259327},
  url       = {https://doi.org/10.1126/science.1259327},
  publisher = {American Association for the Advancement of Science}
}

@article{pawlak2016probing,
  title     = {Probing Atomic Structure and Majorana Wavefunctions in Mono-Atomic {Fe} Chains on Superconducting {Pb} Surface},
  author    = {Pawlak, R{\'e}my and Kisiel, Marcin and Klinovaja, Jelena and Meier, Tobias and Kawai, Shigeki and Glatzel, Thilo and Loss, Daniel and Meyer, Ernst},
  journal   = {npj Quantum Information},
  volume    = {2},
  number    = {1},
  pages     = {16035},
  year      = {2016},
  doi       = {10.1038/npjqi.2016.35},
  url       = {https://doi.org/10.1038/npjqi.2016.35},
  publisher = {Nature Publishing Group}
}

@article{jiang2011majorana,
  title     = {Majorana Fermions in Equilibrium and in Driven Cold-Atom Quantum Wires},
  author    = {Jiang, Liang and Kitagawa, Takuya and Alicea, Jason and Akhmerov, A. R. and Pekker, David and Refael, Gil and Cirac, J. Ignacio and Demler, Eugene and Lukin, Mikhail D. and Zoller, Peter},
  journal   = {Physical Review Letters},
  volume    = {106},
  number    = {22},
  pages     = {220402},
  year      = {2011},
  doi       = {10.1103/PhysRevLett.106.220402},
  url       = {https://doi.org/10.1103/PhysRevLett.106.220402},
  publisher = {American Physical Society}
}

@article{gangadharaiah2011majorana,
  title     = {Majorana Edge States in Interacting One-Dimensional Systems},
  author    = {Gangadharaiah, Suhas and Braunecker, Bernd and Simon, Pascal and Loss, Daniel},
  journal   = {Physical Review Letters},
  volume    = {107},
  number    = {3},
  pages     = {036801},
  year      = {2011},
  doi       = {10.1103/PhysRevLett.107.036801},
  url       = {https://doi.org/10.1103/PhysRevLett.107.036801},
  publisher = {American Physical Society}
}

@article{nunziante2024topological,
  title     = {Topological Phase Diagram of an Interacting {Kitaev} Chain: Mean Field versus {DMRG} Study},
  author    = {Nunziante, Giovanni and Maiellaro, Alfonso and Guarcello, Claudio and Citro, Roberta},
  journal   = {Condensed Matter},
  volume    = {9},
  number    = {1},
  pages     = {20},
  year      = {2024},
  doi       = {10.3390/condmat9010020},
  url       = {https://doi.org/10.3390/condmat9010020},
  publisher = {MDPI}
}

@article{schulz1980critical,
  title     = {Critical Behavior of Commensurate-Incommensurate Phase Transitions in Two Dimensions},
  author    = {Schulz, H. J.},
  journal   = {Physical Review B},
  volume    = {22},
  number    = {11},
  pages     = {5274--5277},
  year      = {1980},
  doi       = {10.1103/PhysRevB.22.5274},
  url       = {https://doi.org/10.1103/PhysRevB.22.5274},
  publisher = {American Physical Society}
}

@article{haldane1981luttinger,
  title     = {{'Luttinger Liquid Theory'} of One-Dimensional Quantum Fluids. I. Properties of the Luttinger Model and Their Extension to the General {1D} Interacting Spinless Fermi Gas},
  author    = {Haldane, F. D. M.},
  journal   = {Journal of Physics C: Solid State Physics},
  volume    = {14},
  number    = {19},
  pages     = {2585--2609},
  year      = {1981},
  doi       = {10.1088/0022-3719/14/19/010},
  url       = {https://doi.org/10.1088/0022-3719/14/19/010},
  publisher = {IOP Publishing}
}

@article{takayoshi2018topological,
  title     = {Topological Transition between Competing Orders in Quantum Spin Chains},
  author    = {Takayoshi, Shintaro and Furuya, Shunsuke C. and Giamarchi, Thierry},
  journal   = {Physical Review B},
  volume    = {98},
  number    = {18},
  pages     = {184429},
  year      = {2018},
  doi       = {10.1103/PhysRevB.98.184429},
  url       = {https://doi.org/10.1103/PhysRevB.98.184429},
  publisher = {American Physical Society}
}

@article{giamarchi1988theory,
  title     = {Theory of Spin-Anisotropic Electron-Electron Interactions in Quasi-One-Dimensional Metals},
  author    = {Giamarchi, T. and Schulz, H. J.},
  journal   = {Journal de Physique},
  volume    = {49},
  number    = {5},
  pages     = {819--835},
  year      = {1988},
  doi       = {10.1051/jphys:01988004905081900},
  url       = {https://doi.org/10.1051/jphys:01988004905081900},
  publisher = {Les {\'E}ditions de Physique}
}

@article{giamarchi1997mott,
  title     = {Mott Transition in One Dimension},
  author    = {Giamarchi, T.},
  journal   = {Physica B: Condensed Matter},
  volume    = {230--232},
  pages     = {975--980},
  year      = {1997},
  doi       = {10.1016/S0921-4526(96)00768-5},
  url       = {https://doi.org/10.1016/S0921-4526(96)00768-5},
  publisher = {Elsevier B.V.}
}

@article{chua2020majorana,
  title     = {Majorana Zero Modes and Their Bosonization},
  author    = {Chua, Victor and Laubscher, Katharina and Klinovaja, Jelena and Loss, Daniel},
  journal   = {Physical Review B},
  volume    = {102},
  number    = {15},
  pages     = {155416},
  year      = {2020},
  doi       = {10.1103/PhysRevB.102.155416},
  url       = {https://doi.org/10.1103/PhysRevB.102.155416},
  publisher = {American Physical Society}
}

@incollection{baxter1985exactly,
  title     = {Exactly Solved Models in Statistical Mechanics},
  author    = {Baxter, Rodney J.},
  booktitle = {Integrable Systems in Statistical Mechanics},
  editor    = {D'Ariano, G. and Montorsi, A. and Rasetti, M.},
  series    = {Series on Advances in Statistical Mechanics},
  volume    = {1},
  pages     = {5--63},
  year      = {1985},
  doi       = {10.1142/9789814415255_0002},
  url       = {https://doi.org/10.1142/9789814415255_0002},
  publisher = {World Scientific},
  address   = {Singapore}
}

@article{mahyaeh2020study,
  title     = {Study of the Phase Diagram of the {Kitaev-Hubbard} Chain},
  author    = {Mahyaeh, Iman and Ardonne, Eddy},
  journal   = {Physical Review B},
  volume    = {101},
  number    = {8},
  pages     = {085125},
  year      = {2020},
  doi       = {10.1103/PhysRevB.101.085125},
  url       = {https://doi.org/10.1103/PhysRevB.101.085125},
  publisher = {American Physical Society}
}

@article{miao2017exact,
  title     = {Exact Solution for the Interacting {Kitaev} Chain at the Symmetric Point},
  author    = {Miao, Jian-Jian and Jin, Hui-Ke and Zhang, Fu-Chun and Zhou, Yi},
  journal   = {Physical Review Letters},
  volume    = {118},
  number    = {26},
  pages     = {267701},
  year      = {2017},
  doi       = {10.1103/PhysRevLett.118.267701},
  url       = {https://doi.org/10.1103/PhysRevLett.118.267701},
  publisher = {American Physical Society}
}

@article{fabrizio1995interacting,
  title     = {Interacting One-Dimensional Electron Gas with Open Boundaries},
  author    = {Fabrizio, Michele and Gogolin, Alexander O.},
  journal   = {Physical Review B},
  volume    = {51},
  number    = {24},
  pages     = {17827},
  year      = {1995},
  doi       = {10.1103/PhysRevB.51.17827},
  url       = {https://doi.org/10.1103/PhysRevB.51.17827},
  publisher = {American Physical Society}
}

@article{pollmann2010entanglement,
  title     = {Entanglement Spectrum of a Topological Phase in One Dimension},
  author    = {Pollmann, Frank and Turner, Ari M. and Berg, Erez and Oshikawa, Masaki},
  journal   = {Physical Review B},
  volume    = {81},
  number    = {6},
  pages     = {064439},
  year      = {2010},
  doi       = {10.1103/PhysRevB.81.064439},
  url       = {https://doi.org/10.1103/PhysRevB.81.064439},
  publisher = {American Physical Society}
}

@article{baxter1971eight,
  title     = {Eight-Vertex Model in Lattice Statistics},
  author    = {Baxter, Rodney J.},
  journal   = {Physical Review Letters},
  volume    = {26},
  number    = {14},
  pages     = {832--833},
  year      = {1971},
  doi       = {10.1103/PhysRevLett.26.832},
  url       = {https://doi.org/10.1103/PhysRevLett.26.832},
  publisher = {American Physical Society}
}

@article{fidkowski2010effects,
  title     = {Effects of Interactions on the Topological Classification of Free Fermion Systems},
  author    = {Fidkowski, Lukasz and Kitaev, Alexei},
  journal   = {Physical Review B},
  volume    = {81},
  number    = {13},
  pages     = {134509},
  year      = {2010},
  doi       = {10.1103/PhysRevB.81.134509},
  url       = {https://doi.org/10.1103/PhysRevB.81.134509},
  publisher = {American Physical Society}
}

@article{fidkowski2011topological,
  title     = {Topological Phases of Fermions in One Dimension},
  author    = {Fidkowski, Lukasz and Kitaev, Alexei},
  journal   = {Physical Review B},
  volume    = {83},
  number    = {7},
  pages     = {075103},
  year      = {2011},
  doi       = {10.1103/PhysRevB.83.075103},
  url       = {https://doi.org/10.1103/PhysRevB.83.075103},
  publisher = {American Physical Society}
}

@article{bortolin2019quantum,
  title={Quantum phase diagram of Shiba impurities from bosonization},
  author={Bortolin, Tom{\'a}s and Iucci, An{\'\i}bal and Lobos, Alejandro M},
  journal={arXiv preprint arXiv:1904.02834},
  year={2019}
}

@article{suleiman2026electrostatically,
  title={Electrostatically induced topological phase transitions in polyacetylene molecules},
  author={Suleiman, Tom{\'a}s and Iucci, An{\'\i}bal and Lobos, Alejandro M},
  journal={Physical Review B},
  volume={114},
  number={6},
  pages={065425},
  year={2026},
  publisher={APS}
}

\end{document}